\documentclass[twocolumn, twocolappendix]{aastex631}
\usepackage{enumitem}

\shorttitle{CME-CME interaction}
\shortauthors{Mayank et al.}

\begin{document}

\title{Study of Evolution and Geo-effectiveness of CME-CME Interactions using MHD Simulations with SWASTi framework}

\correspondingauthor{Prateek Mayank}
\email{prateekmayank9@gmail.com}

\author[0000-0001-8265-6254]{Prateek Mayank}
\affiliation{Department of Astronomy, Astrophysics and Space Engineering, Indian Institute of Technology Indore, India}

\author[0000-0002-1037-348X]{Stefan Lotz}
\affiliation{South African National Space Agency, Hermanus, South Africa}

\author[0000-0001-5424-0059]{Bhargav Vaidya}
\affiliation{Department of Astronomy, Astrophysics and Space Engineering, Indian Institute of Technology Indore, India}
\affiliation{Center of Excellence in Space Sciences India, IISER Kolkata, India}

\author[0000-0003-2740-2280]{Wageesh Mishra}
\affiliation{Indian Institute of Astrophysics, Bangalore, India}

\author[0000-0003-2693-5325]{D. Chakrabarty}
\affiliation{Space and Atmospheric Sciences Division, Physical Research Laboratory, Ahmedabad, India}


\begin{abstract}

The geo-effectiveness of Coronal Mass Ejections (CMEs) is a critical area of study in space weather, particularly in the lesser-explored domain of CME-CME interactions and their geomagnetic consequences. This study leverages the SWASTi framework to perform 3D MHD simulation of a range of CME-CME interaction scenarios within realistic solar wind conditions. The focus is on the dynamics of the initial magnetic flux, speed, density, and tilt of CMEs, and their individual and combined impacts on the disturbance storm time (Dst) index. Additionally, the kinematic, magnetic, and structural impacts on the leading CME, as well as the mixing of both CMEs, are analyzed. Time series in-situ studies are conducted through virtual spacecraft positioned along three different longitudes at 1 AU. Our findings reveal that CME-CME interactions are non-uniform along different longitudes due to the inhomogeneous ambient solar wind conditions. A significant increase in the momentum and kinetic energy of the leading CME is observed due to collisions with the trailing CME, along with the formation of reverse shocks in cases of strong interaction. These reverse shocks lead to complex wave patterns inside CME2, which can prolong the storm recovery phase. Furthermore, we observed that the minimum Dst value decreases with an increase in the initial density, tilt, and speed of the trailing CME.

\end{abstract}

. 

\section{Introduction} \label{sec:intro}

Coronal Mass Ejections (CMEs) are significant drivers of space weather, characterized by the ejection of massive amounts of magnetized plasma from the Sun's corona. When multiple CMEs are launched in quick succession, their interactions—termed CME-CME interactions—can dramatically enhance their space weather impact. CME-CME interactions occur when a faster CME overtakes a slower one, leading to a complex interplay of their shocks, magnetic fields and plasma structures. Studying these interactions is crucial as they can significantly amplify geomagnetic storms, particle acceleration, and other space weather phenomena, posing greater risks to technological systems and human activities \citep{gopalswamy_2001_radio, lugaz_2017_the, scolini_2020_cmecme}.

The complexity of CME-CME interactions arises from the already intricate dynamics present in individual CME and solar wind (SW) interactions. In single CME-SW interactions, ambient solar wind conditions can significantly modify the CME's trajectory, speed, internal properties, and structure \citep{temmer_2011_influence, shen_2012_acceleration, wu_2016_numerical, prateekmayank_2023_swasticme}. When a trailing CME catches up and collides with a leading CME, resulting in a CME-CME-SW interaction, the scenario becomes even more complex. Studies have reported a wide range of collision types between CMEs, from inelastic \citep{mishra_2014_evolution}, nearly elastic \citep{mishra_2015_kinematics}, superelastic collisions \citep{shen_2012_superelastic}, to merging-like processes \citep{temmer_2012_characteristics}.

Multiple observational and simulation studies have made significant progress in understanding the evolution of CMEs during interactions. \cite{shen_2016_turn} showed that the final speeds depend on the relative masses of the CMEs as well as their relative speeds. Through 2.5D simulations, \cite{poedts_2003} noted that the acceleration of the leading CME increases as the mass of the trailing CME increases. In addition to speed, CME-CME interactions can also lead to deflection of the CMEs \citep{lugaz_2012_the, shen_2012_superelastic}. Observational and simulation-based studies have also demonstrated that the expansion of the radial width of the leading CME decreases as the trailing CME impacts and compresses its rear \citep{nolugaz_2005_numerical, xiong_2006_magnetohydrodynamic}.

In addition to changes in CME properties, multiple studies have elucidated the behavior of shocks within magnetic clouds. \cite{mvandas_1997_mhd} highlighted that shocks propagate more swiftly inside magnetic clouds due to enhanced fast magnetosonic speeds, potentially leading to shock-shock merging near the cloud’s nose while maintaining distinct shocks at the flanks. Other numerical studies have also reported that weak or slow shocks within regions of elevated magnetosonic speeds dissipate inside the magnetic cloud \citep{xiong_2006_magnetohydrodynamic, lugaz_2007_numerical}. Additionally, \cite{farrugia_2004_evolutionary} reported the merging or dissipation of shocks through Helios and ISEE-3 measurements, showing a decrease from four shocks at 0.67 AU to two at 1 AU. \cite{nolugaz_2005_numerical} provided a comprehensive analysis, delineating four primary phases of shock property changes during such interactions.

Several studies have observed that CME-CME interactions are a common source of double-dip and multiple-dip geomagnetic storms \citep{zhang_2008_interplanetary, richardson_2008_multiplestep}. Much of the understanding about the impact of CME's initial properties -- that is, right after eruption, measured at around 0.1 AU -- on their geo-effectiveness has come from MHD simulations. \cite{scolini_2020_cmecme} quantified the impact of interactions on the geo-effectiveness of individual CMEs using the European heliospheric forecasting information asset \citep[EUHFORIA, ][]{Pomoell2018} spheromak CME model. They found that the time interval between the CME eruptions and their relative speeds are critical factors in determining the resulting impact of the CME-CME structure. Additionally, \cite{koehn_2022_successive} conducted MHD simulations of spheromak CMEs with a uniform outflowing solar wind and found that the orientation and handedness of a given CME can significantly impact the conservation and loss of magnetic flux in the CME.

Although several observational studies have shown the consequences of CME-CME interactions, they have not been very successful in elucidating the interaction process itself. Most studies have primarily focused on different aspects of the interaction without attempting to explore a global view. While multiple numerical studies have provided great insights into these interaction processes, particularly shock evolution, there have been very few studies on CME-CME interactions occurring within realistic dynamic ambient solar wind conditions \citep{scolini_2020_cmecme}. Given the complexities and limitations in observational and simplified simulation studies, MHD ensemble simulations with realistic solar wind backgrounds offer a powerful tool to obtain a global view.

In this work, we used the Space Weather Adaptive SimulaTion framework \citep[SWASTi;][]{mayank_2022_swastisw, prateekmayank_2023_swasticme} to conduct ensemble MHD simulations with a data-driven solar wind background. Our aim is to identify global trends and understand the impact of initial CME properties on the geo-effectiveness of the CME-CME structure. We used the SWASTi-CME module to simulate two-successive flux rope CMEs and trace their evolution in the inner heliosphere. Further, we quantified their geo-effectiveness using the empirical Dst relation given by \cite{obrien_2000_forecasting}.

The rest of the paper is organised as follows: Section \ref{sec:models} provides a brief description of the numerical models for solar wind, CME, and Dst index used in this work. In Section \ref{sec:cases}, we describe the ensemble cases with initial values of CME and an overview of the CME-CME-SW interaction scenario. Section \ref{sec:evolution} contains the ensemble simulation results and detailed analysis of the evolution of shock and leading CME properties. Further, Section \ref{sec:geo-effectiveness} presents a statistical analysis of the variations in the minimum and cumulative Dst indices. Finally, Section \ref{sec:summary} provides the discussion and conclusions of the work.

\section{Numerical Models} \label{sec:models}
To perform the ensemble study of CME-CME interactions within a realistic solar wind background, we utilized the SWASTi framework. The three-dimensional physics-based models for solar wind and CMEs are described in the following subsections. Additionally, an empirical Dst relation, based on in-situ plasma properties, has been employed to analyze the geo-effectiveness of these energetic events. The relevant equations and comparisons with some specific events are presented in subsequent subsections.

\subsection{Solar Wind Model} \label{sec: SW_model}

In order to simulate the background solar wind, we used observation based input from GONG magnetogram, which provides the magnetic field at solar surface. The fieldlines are then extrapolated to source surface using the  potential field source surface \citep[PFSS;][]{altschuler_1969_magnetic} technique. Based on this extrapolation, the solar wind speed at the initial boundary of the MHD domain ($V_{in}$), located at 0.1 AU, is determined using a modified version of the Wang-Sheeley-Arge relation \citep[WSA;][]{arge_2003_improved}:

    \begin{equation}\label{eq:WSA}
       V_{{in}} = v_{1} + \frac{v_{2}}{(1+f_s)^{\frac{2}{9}}} \times \Bigg(1.0 - 0.8\,exp \Bigg(-\bigg(\frac{d}{w}\bigg)^{\beta}\Bigg)\Bigg)^{3}
   \end{equation}

where,  $v_{1}, v_{2},$ and $\beta$ are parameters set at 250 km s$^{-1}$, 675 km s$^{-1}$, and 1.25, respectively. $f_s$ represents the areal expansion factor of the flux tube, $d$ is the minimum angular separation of the foot-point from the coronal hole boundary, and $w$ is the median of $d$ for field lines that extend to Earth's location. The initial density at 0.1 AU was estimated by equating the kinetic energy due to obtained WSA speed with that of the fast solar wind. Here we have assumed fast wind parameters as  $n_{\rm fsw}$ = 200 cm$^{-3}$ and speed $v_{\rm fsw}$ = 600 km s$^{-1}$. The temperature was determined based on constant thermal pressure of 6.0 nPa. The magnetic field was derived from a velocity-dependent empirical relation. For detailed methodology, readers can refer to Section 2 of \cite{mayank_2022_swastisw}. In this study, magnetic field strength ($B_{\rm fsw}$) associated with 650 km s$^{-1}$ is assumed to be 300 nT at the MHD domain's initial boundary.

After setting the necessary parameters at 0.1 AU based on the semi-empirical coronal model, the time-dependent 3D ideal MHD equations were solved using the PLUTO code \citep{mignone_2007_pluto}. Computations were conducted on a uniform static grid in spherical coordinates, employing a finite volume method for the simulation. The set of conservative equations used in the MHD simulations is outlined in \cite{mayank_2022_swastisw}, with a specific heat ratio of 1.5 for to the solar wind plasma. The computational domain for the heliosphere extended from 0.1 AU to 2.1 AU radially ($r$), $\pm$60\textdegree{} azimuthally ($\theta$), and 0\textdegree{} to 360\textdegree{} meridionally ($\phi$), structured on a grid resolution of $512 \times 61 \times 181$.

\subsection{CME Model} \label{sec: CME_model}
To simulate magnetized CMEs, we employed the CME module of SWASTi framework \citep{prateekmayank_2023_swasticme}, which is based on the Flux Rope in 3D \citep[FRi3D;][]{isavnin_2016_fried} model. This model incorporates the three-dimensional magnetic field configuration of a CME and accounts for major deformations to accurately reconstruct its global geometrical shape. In this study, the FRi3D model was used to construct the 3D magnetized shell of the CME at 0.1 AU, serving as the initial state for the MHD domain. The CME geometry forms a classic croissant-like shape anchored at both ends to the Sun in the beginning and then cut for non-hindrance eruption of the trailing CME.

For single flux rope CMEs, cutting the legs when their speed matches the ambient solar wind is commonly used to ensure smooth integration with the background. However, in CME-CME interactions, this timing becomes more complex, as the trailing CME may erupt before the leading CME legs have slowed enough for cutting. Additionally, this method does not ensure a consistent CME structure across different cases with varying inner-boundary conditions, which is critical for this study. To address these issues, we implemented a fixed-duration insertion process for all simulation cases, where the duration was optimized to ensure that the average CME leg speed closely matches the ambient solar wind. This approach maintains structural consistency across all simulations and minimizes disruptions in the solar wind outflow from the inner boundary.

Initially, the cross-section of the CME is assumed to be circular, with the radius varying in proportion to the heliocentric distance. The CME structure is populated with magnetic field lines that have a constant twist of two. In this work, the center of the CME footpoints was set at 0\textdegree\ latitude and 0\textdegree\ longitude. All CMEs had the following fixed parameters: flattening = 0.5, pancaking = 0.6, chirality = -1, polarity = -1, half-width = 45\textdegree, and half-height = 22.5\textdegree. Other properties were varied in the ensemble study, with their exact values detailed in Section \ref{sec:cases}.

Once the FRi3D CME structure is formed with its leading edge at 0.1 AU, it is allowed to evolve in the MHD domain. The process of integrating the CME into the MHD domain involves gradually updating the boundary conditions to match the evolving CME structure, ensuring a smooth transition and accurate representation of the CME's impact on the surrounding solar wind. Overall, the use of flux rope CME model in this study provides a detailed and realistic simulation of CME-CME interactions, allowing for a comprehensive analysis of their evolution, and the influence of initial conditions on geo-effectiveness.

\subsection{Dst Estimation} \label{sec: Dst_model}
We use the disturbance storm time (Dst) index to quantify the geoeffectiveness of the simulated CME plasma cloud. The Dst index estimates the storm time ring current strength without the influence of magnetopause and quiet time ring currents \citep{obrien_2000_forecasting}. There are many other indices that may be used to describe the coupling between solar wind and magnetosphere \citep[see][for a good summary]{lockwood_solar_2022} but since we are most interested in the kind of intense events that would arise from the effect of merged CMEs on the geomagnetic field, we feel that Dst is an appropriate proxy for this study.

To quantify the geoeffectiveness of the CME-CME structure upon its arrival at 1 AU, we positioned virtual spacecraft within the simulation domain at 0\textdegree\ (along Sun-Earth line) and $\pm$10\textdegree\ longitudes. These virtual spacecraft recorded the plasma properties in real-time with a 5-minute cadence. Based on these in-situ properties, the Dst indices at each time steps were computed using the empirical equations provided by \cite{obrien_2000_forecasting}. Further details about the equations used and their comparison with observed events are presented in Section \ref{sec: Dst_relation}.

\section{Simulation Cases} \label{sec:cases}

    \begin{table*}
    \begin{minipage}{\textwidth}
    \centering
    \caption{CME initial properties of all the ensemble cases.}
    \label{tab:all_cases}
    \begin{tabular}{lcrrrr}
    \textbf{Case No.} & \textbf{Name} & \textbf{\begin{tabular}[c]{@{}c@{}}Vel\_t\\ ($10^{3} kms^{-1}$)\end{tabular}} & \textbf{\begin{tabular}[c]{@{}c@{}}Density\\ ($10^{-18} kgm^{-3}$)\end{tabular}} & \textbf{\begin{tabular}[c]{@{}c@{}}Magnetic Flux\\ ($10^{12} Wb$)\end{tabular}} & \multicolumn{1}{c}{\textbf{\begin{tabular}[c]{@{}c@{}}Tilt\\ (degree)\end{tabular}}} \\ \hline
    Case 0 & Single CME & v$_t$1 = 0.8 & $\rho$1 = 3 & $\Phi_B$1 = 7 & $\tau1$ = 0 \\ \hline
    Case 1 & LSLDLF0 & v$_t$2 = 1 & $\rho$2 = 1 & $\Phi_B$2 = 5 & $\tau2$ = 0 \\
    Case 2 & LSLDLF1 & v$_t$2 = 1 & $\rho$2 = 1 & $\Phi_B$2 = 5 & $\tau2$ = 45 \\
    Case 3 & LSLDHF0 & v$_t$2 = 1 & $\rho$2 = 1 & $\Phi_B$2 = 9 & $\tau2$ = 0 \\
    Case 4 & LSLDHF1 & v$_t$2 = 1 & $\rho$2 = 1 & $\Phi_B$2 = 9 & $\tau2$ = 45 \\
    Case 5 & LSHDLF0 & v$_t$2 = 1 & $\rho$2 = 5 & $\Phi_B$2 = 5 & $\tau2$ = 0 \\
    Case 6 & LSHDLF1 & v$_t$2 = 1 & $\rho$2 = 5 & $\Phi_B$2 = 5 & $\tau2$ = 45 \\
    Case 7 & LSHDHF0 & v$_t$2 = 1 & $\rho$2 = 5 & $\Phi_B$2 = 9 & $\tau2$ = 0 \\
    Case 8 & LSHDHF1 & v$_t$2 = 1 & $\rho$2 = 5 & $\Phi_B$2 = 9 & $\tau2$ = 45 \\
    Case 9 & HSLDLF0 & v$_t$2 = 1.1 & $\rho$2 = 1 & $\Phi_B$2 = 5 & $\tau2$ = 0 \\
    Case 10 & HSLDLF1 & v$_t$2 = 1.1 & $\rho$2 = 1 & $\Phi_B$2 = 5 & $\tau2$ = 45 \\
    Case 11 & HSLDHF0 & v$_t$2 = 1.1 & $\rho$2 = 1 & $\Phi_B$2 = 9 & $\tau2$ = 0 \\
    Case 12 & HSLDHF1 & v$_t$2 = 1.1 & $\rho$2 = 1 & $\Phi_B$2 = 9 & $\tau2$ = 45 \\
    Case 13 & HSHDLF0 & v$_t$2 = 1.1 & $\rho$2 = 5 & $\Phi_B$2 = 5 & $\tau2$ = 0 \\
    Case 14 & HSHDLF1 & v$_t$2 = 1.1 & $\rho$2 = 5 & $\Phi_B$2 = 5 & $\tau2$ = 45 \\
    Case 15 & HSHDHF0 & v$_t$2 = 1.1 & $\rho$2 = 5 & $\Phi_B$2 = 9 & $\tau2$ = 0 \\
    Case 16 & HSHDHF1 & v$_t$2 = 1.1 & $\rho$2 = 5 & $\Phi_B$2 = 9 & $\tau2$ = 45 \\ \hline      
    \end{tabular}
    \end{minipage}
    \end{table*}

To investigate the evolution and geo-effectiveness of CME-CME interaction events, we conducted a series of simulations involving two interacting CMEs, with second CME having varying attributes. The primary objective was to understand how alterations in initial coronal properties --- such as speed, density, and magnetic flux --- affect their interaction, in-situ properties and the Dst index at 1AU. This focus is driven by the understanding that in-situ speed, density, and the southward magnetic field component play a pivotal role in determining the intensity of geomagnetic storms. Therefore, assessing the effects of these initial conditions at 1 AU is crucial. Moreover, recent research highlighting the significance of CME tilt in heliospheric evolution \citep{prateekmayank_2023_swasticme, lugaz_2013_the} prompted us to also examine the impact of relative tilt between interacting CMEs.

For this ensemble simulation, we selected the background solar wind conditions from the Carrington rotation (CR) 2270 period, corresponding to April 2023. This period was particularly significant due to an intense geomagnetic storm caused by a halo CME eruption. The Dst values of this CME have been discussed in Section \ref{sec: Dst_relation}, where the empirical Dst is compared with the observed values. In the simulation, the first CME (hereafter referred as CME1) was injected at the inner-boundary (0.1 AU) on 2023-05-12 3:00, with the second CME (hereafter referred as CME2) following 25 hours later. This schedule was optimised to ensure that the CME2 reaches 0.1 AU after the complete insertion of CME1 in the computational domain and they have significant interaction before reaching 1 AU.

Several studies have highlighted the influence of ambient solar wind on CME evolution \citep{temmer_2011_influence, wu_2016_numerical, prateekmayank_2023_swasticme}. To ascertain if similar impacts are present in CME-CME interactions, this study analyzes in-situ profiles at three different longitudinal positions: 0\textdegree \,and $\pm$10\textdegree. Here, 0\textdegree\ corresponds to the Sun-Earth line, while -10\textdegree\ and +10\textdegree\ represent the eastern and western sides of the solar disk, respectively. Additionally, the projected trajectory allows the CME1 to interact with a stream interaction region (SIR), thereby enabling a detailed study of the potential impact of inhomogeneity in ambient solar wind. 

To determine the optimal number of cases for the ensemble study, our objective was to select a sufficient number of cases to identify the trends in the properties while ensuring each case could be thoroughly analyzed. To examine the effects of speed, density, and magnetic flux, we varied these parameters for the CME2 while keeping the CME1's values constant. Specifically, for the CME2, we employed two sets of density and magnetic flux values, one set lower and another higher than the CME1’s values. Furthermore, we chose two velocity values for the CME2, both higher than the first’s and increasing incrementally, to guarantee their interaction before 1 AU. Regarding the relative tilt, two tilt angles were applied to the CME2 while the tilt of CME1 was kept constant. The overarching approach was to maintain consistent properties for the CME1 across all cases, thereby enabling a direct comparison with single CME1 simulation.

The specific values of the properties for the various cases are presented in Table 1. With two values for each of the four properties, the total number of cases in this ensemble study amounts to 16. The density of the CMEs was set of the order of $10^{-18}$ $\rm kg m^{-3}$ \citep{temmer_2021_deriving}, while the magnetic flux values were between $10^{12}$ - $10^{13}$ $\rm Wb$ \citep{scolini_2020_cmecme}. The CME apex speed values ranged approximately from 1300 to 1500 $\rm km s^{-1}$, and the tilt values were 0\textdegree\, and 45\textdegree. To facilitate clear and easy reference during discussion, each case has been assigned a specific nomenclature.  This naming convention consists of four elements: the first two letters denote low speed (LS)/high speed (HS) CME, followed by a pair of letters for low/high density and magnetic flux, respectively. The final character in each case name denotes the tilt, with `0' representing no tilt and `1' indicating a 45\textdegree \, tilt. For example, the name `HSHDLF1' corresponds to a case with higher speed, higher density, lower magnetic flux, and a 45\textdegree \, tilt of CME2 with respect to first. Additionally, a simulation of just the CME1 was conducted to serve as a basis for comparing all interaction cases, thereby highlighting the impact of such interactions on the front CME.

     \begin{figure*}
           \centering
           \includegraphics[width = \textwidth]{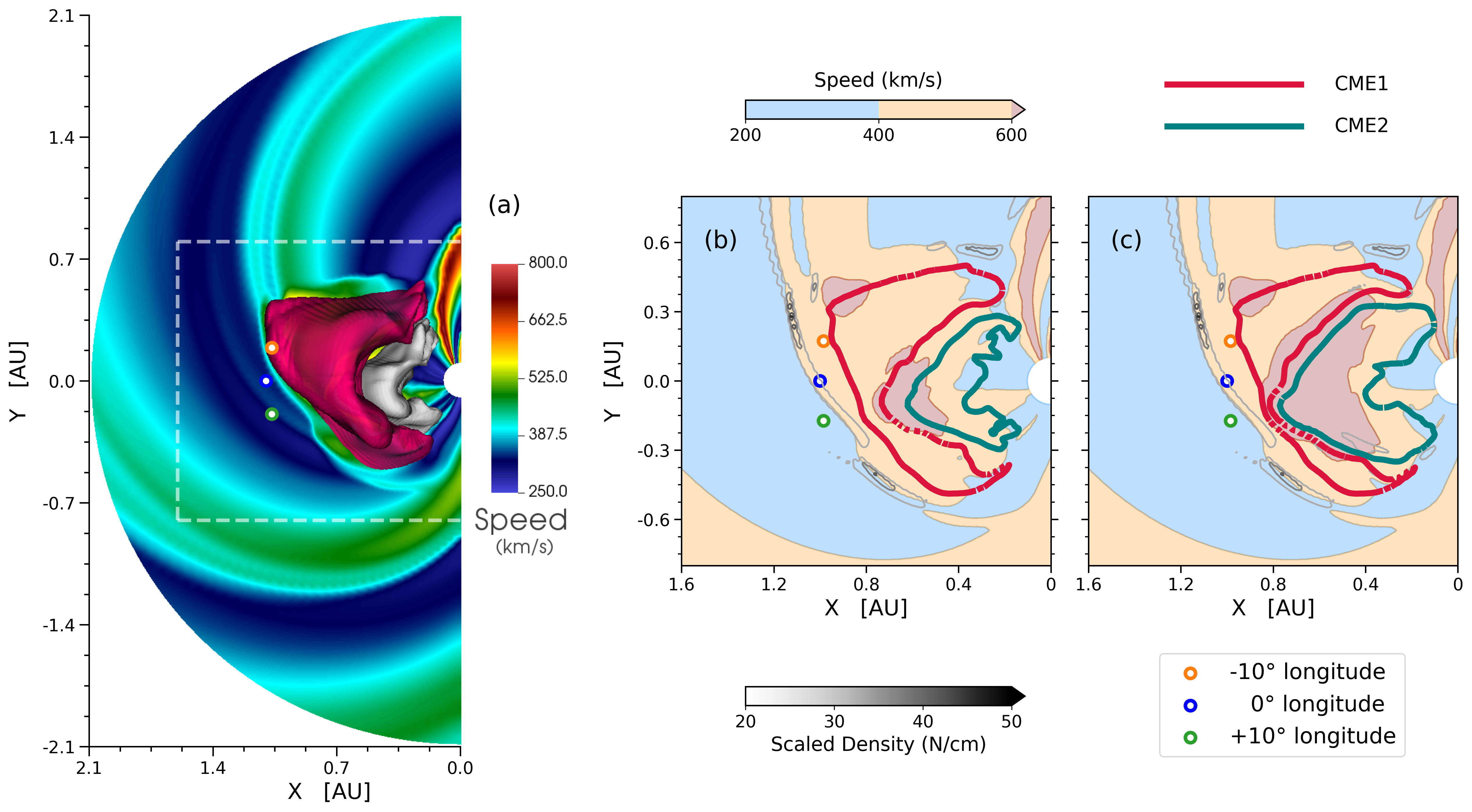}
           \caption{The subplot (a) displays a 3D view of the leading (crimson) and trailing (white) structures of CMEs, overlaid on a 2D snapshot of solar wind speed in the equatorial plane. The dashed white box indicates the region shown in subplots (b) and (c). They depict the traced CME boundaries for cases (b) LSLDLF0 and (c) LSHDLF0, with color-filled contours illustrating plasma speed regions: slow, fast, and $>$600 km/s. Grey line contours indicate high-density regions, with darker shades representing higher scaled densities. The blue, green, and orange dots mark the positions of virtual spacecraft at -10\textdegree, 0\textdegree and +10\textdegree.}
           \label{fig:2D_example}
    \end{figure*}

 Figure \ref{fig:2D_example} (a) presents a snapshot capturing the evolution of two interacting CMEs in the inner heliosphere, corresponding to the LSLDLF0 case as CME1 reaches 1 AU. It reveals 2D cross-section of the speed profile along the $r - \phi$ plane at 0\textdegree \, latitude, alongside 3D isosurfaces of the two CMEs. The background solar wind speed is depicted using a color map, with CME1 outlined in crimson and CME2 in a white hue. This image also illustrates the asymmetric expansion of CME1's leading edge, shaped by the variable speeds of the ambient solar wind streams.

Subplot \ref{fig:2D_example}(b) provides a detailed cut-out from \ref{fig:2D_example}(a), while subplot \ref{fig:2D_example}(c) is akin to \ref{fig:2D_example}(b) but represents the LSHDLF0 case. The red and teal colored contours represent the traced boundaries in the equatorial plane of the 3D CME structures shown in \ref{fig:2D_example}(a). These subplots showcase the filled contours for speed, as well as line contours for scaled density. The three dots at 1 AU mark the positions of virtual spacecrafts measuring in-situ plasma properties at 0\textdegree \,and $\pm$10\textdegree \, longitudes. Notably, there is a fast stream at $\phi = -10$\textdegree \,, a slow-speed stream at $\phi = +10$\textdegree \,, with $\phi = 0$\textdegree \,  location approximately at the juncture of these streams. Due to which, the disparity in the expansion of CME1's flanks is apparent along these longitudes, with the top flank showing more pronounced expansion compared to the bottom flank. The SIR has higher density, as compared to fast or slow streams, making them more effective in compressing CME1 when CME2 approaches from behind, as depicted in subplot \ref{fig:2D_example}(c). Here, the intensified compression along -10\textdegree \, longitude is clearly visible.

It is important to emphasize that dark orange regions of high speed do not necessarily represent shock structures but rather highlight areas where the speed exceeds 600 km/s. Comparing subplots \ref{fig:2D_example}(b) and \ref{fig:2D_example}(c), the $>$600 km/s region in (c) extends significantly further in both radial and longitudinal directions. Specifically, this high-speed region extends about 0.3 AU radially and covers approximately 30\textdegree\, in (b), while in (c), it stretches to around 0.5 AU radially and spans roughly 75\textdegree, indicating a stronger shock associated with CME2 in the high-density case. Additionally, the increased radial width across all longitudes in (c) suggests that CME2 has expanded more, leading to greater compression of CME1. A more detailed analysis on the evolution of the CMEs in the heliosphere is discussed in the subsequent section.

\section{Evolution in Heliosphere} \label{sec:evolution}
One of the key aspects of CME-CME interaction is the change in their properties as they traverse the inner heliosphere. The nature and extent of this interaction play a critical role in determining their characteristics upon reaching Earth. These alterations directly affect space weather forecasting, underscoring the importance of understanding these dynamic processes. Through ensemble 3D MHD simulations with realistic solar wind conditions, we have sought to explore the following dynamics.

\subsection{Shock Dynamics}

The complex process of CME-CME interaction can be  segmented into progressive phases based on the trailing shock associated with CME2. \cite{nolugaz_2005_numerical} identified four distinct stages: 

\begin{enumerate}
    \item the shock propagates through the solar wind before reaching the rear of CME1,
    \item upon impact, the shock advances inside CME1,
    \item subsequently, the shock begins interacting with the sheath of CME1,
    \item finally, the merging of the shocks commences.
\end{enumerate}

All interaction scenarios evolve through these stages, and depending on the properties of the CMEs, the specific stage at which they arrive at 1 AU may vary.

\begin{figure*}
           \centering
           \includegraphics[width = \textwidth]{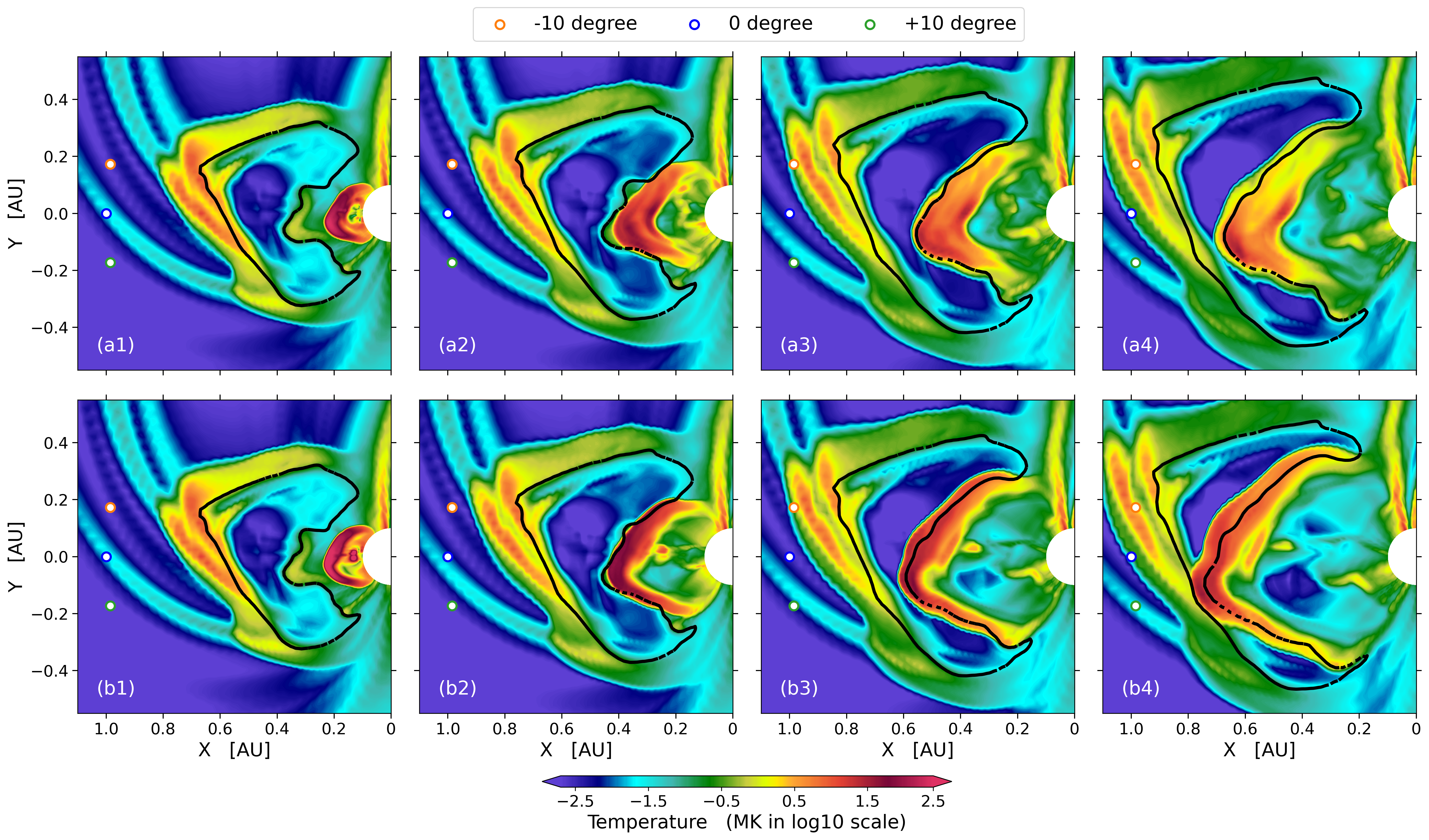}
           \caption{The subplots demonstrates the evolution of trailing shock in the inner-heliosphere for cases LSLDLF0 (a1-a4) and LSHDLF0 (b1-b4). The displayed colormap is corresponding to the plasma temperature in the logarithmic scale.}
           \label{fig:2D_shock}
    \end{figure*}

Figure \ref{fig:2D_shock} demonstrates these evolutionary stages. The subplots display the temperature profile in a logarithmic scale, emphasizing the sharp gradient of the shock associated with CME2. The corresponding scaled density profile is shown in Figure \ref{fig:rho_shock}. Subplots (a1) to (a4) pertain to the LSLDLF0 case, while subplots (b1) to (b4) relate to the LSHDLF0 case, each separated by a time interval of 6.65 hours. The only difference between the upper and lower rows is the density of CME2. Although initially similar, as time progresses, significant differences emerge. The interaction commences earlier in the HD case, with the trailing shock penetrating deeper into the leading magnetic cloud than in the LD case. As CME1 reaches 1 AU, the HD case nearly reaches stage 4, while the LD case has just entered stage 3. Notably, when different longitudes are considered, both cases exhibit varying stages at each location. Thus, depending on the portion of the CME-CME interaction structure encountering Earth, it may be at a different stage of evolution.

\subsubsection{Non-uniform Shock Interaction}
Most of the previous studies \citep[][and references therein]{lugaz_2017_the} have focused on defining a single stage for the entire CME-CME interaction structure. However, given the relatively small scale of Earth's magnetosphere compared to the CME structure, examining the evolution of these stages at different longitudes is critical. For instance, the structures passing through the 1 AU locations marked by three dots in Figure \ref{fig:2D_shock} at 0\textdegree\ and ±10\textdegree\ longitudes are at different stages. The trailing shock does not extend to the sheath of CME1 along the -10\textdegree\ longitude in any of the cases. The radial evolution of this shock is greater along 0\textdegree\ and even more so along +10\textdegree, where it interacts with the sheath and merges with the first shock in some instances. The uneven evolution of these stages is primarily caused by the deformation of CME1 due to the ambient solar wind preceding it, particularly due to the SIR in front of it in these specific situations. This SIR positions the top flank of the CME predominantly within the fast solar wind stream, while the bottom flank experiences a significantly stronger anti-radial drag force. This dynamic results in an overexpansion of the upper flank and an underexpansion of the lower flank, as depicted in Figure \ref{fig:2D_example}.

This 4-stage evolution concept can be applied both globally across the entire structure and locally along different longitudes to study the progression of CME-CME interactions comprehensively. To facilitate a robust comparison of different cases with varying initial properties, it is crucial to identify both global and local stages. For this purpose, we utilized the concept of virtual spacecraft to observe the in-situ plasma conditions at the front and rear of CME1 along three longitudes (0\textdegree, and $\pm$10\textdegree). Six comoving virtual spacecraft were positioned, with three at the rear and three at the front of CME1. Figure \ref{fig:shock_duration} displays the in-situ speed profiles collected by these spacecraft for the HSHDLF0 case. In the top subplot (\ref{fig:shock_duration}a), the solid lines represent data from the rear of CME1, while the dashed lines are from the front. At the rear of CME1, the trailing shock arrives with a time difference of 1 hour between the $\pm$10\textdegree\ longitudes, resulting in a similar speed jump along these longitudes due to the equivalent shock propagation time. However, at the front of CME1, the arrival time difference of the trailing shock between these longitudes is much larger. This is because the radial width of CME1 varies significantly between +10\textdegree\ and -10\textdegree\ longitudes, leading to an enhanced temporal gap in shock arrival. Furthermore, since the radial extent of CME1 is greater along -10\textdegree\, the trailing shock travels a longer distance and loses more energy, resulting in a smaller speed jump compared to +10\textdegree\ longitude.

\subsubsection{Shock Propagation Duration}
    
\begin{figure}
           \centering
           \includegraphics[width = \columnwidth]{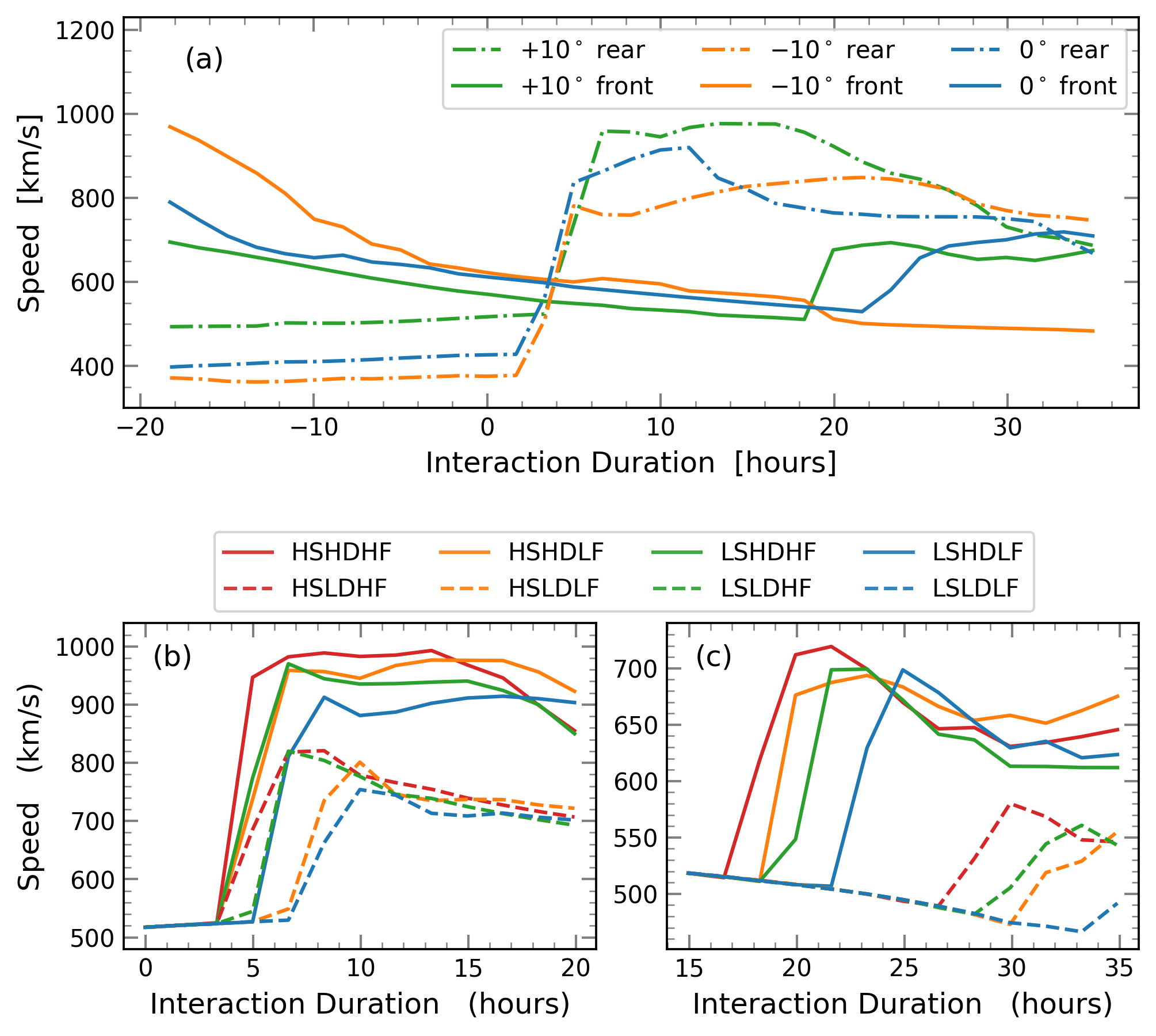}
           \caption{The arrival of trailing shock at the rear and front of CME1. Subplot (a) shows the result for HSHDLF0 case along 0\textdegree\ \& $\pm$10\textdegree. (b) depicts the speed variation at the rear of CME1, along -10\textdegree, for all cases with 0$^\circ$ tilt. Similarly (c) shows the speed at the front of CME1.}
           \label{fig:shock_duration}
    \end{figure}

For comparison across all ensemble cases, we focused on the +10\textdegree\ longitude to study the properties of shock propagation. This approach examines the time it takes for the trailing shock to travel from the rear to the front of CME1, a duration primarily influenced by two factors: the radial width of CME1 and the strength of the trailing shock. Given that the CME1 properties remain consistent across all cases, the strength of the trailing shock emerges as the sole determinant of this duration. The bottom left and right subplots in Figure \ref{fig:shock_duration} illustrate the arrival and departure times of the trailing shock from the CME1 structure, revealing two distinct behaviors influenced by the initial density of CME2. The speed jump due to the shock was consistently higher in all HD cases compared to all LD cases. Although HD cases had similar arrival times, the sharpness of the shock was greatest in the HSHDHF case and least in the LSHDLF case. As the trailing shock reaches the front of CME1, the temporal differences between the cases become more pronounced, with the shock propagation duration for the strongest (HSHDHF0) and weakest (LSLDLF1) shock cases widening from 13.3 to 28.26 hours.

The trailing shock propagation time ($\Delta$) for all ensemble cases is presented in Table 2. The duration is shortest for the HSHDHF case in both with and without relative tilt scenarios, and progressively increases as the initial flux, speed, and density decrease, with the longest duration observed in the LSLDLF case. Cases with a higher initial density consistently exhibit shorter $\Delta$ values. Since this duration is directly correlated with shock strength, it suggests that higher density CMEs generate stronger shocks. Elevated density leads to an increase in internal pressure within the CME, which amplifies the pressure differential with the ambient solar wind. This, in turn, results in an accelerated rate of CME expansion. Moreover, greater momentum allows the CME to plow more effectively through the solar wind plasma, especially in the low-density environment of the preconditioned solar wind. Therefore, a combination of a faster expansion rate and more efficient plowing contributes to the observed trend. Furthermore, cases with 0\textdegree\, tilt consistently show smaller $\Delta$ values compared to those with 45\textdegree\, tilt. This could potentially be due to the preconditioned solar wind: CME2, having no relative tilt, follows the same trajectory as CME1 and encounters less drag force from the low-density solar wind swept by CME1. However, with non-zero relative tilt, the structure of CME2 does not completely align with the cleared path and faces greater resistance from the ambient solar wind, potentially leading to a reduction in shock strength.

\begin{table}
\centering
\caption{Trailing shock propagation time inside CME1 for all ensemble cases.}
\begin{tabular}{cccc}
\multicolumn{1}{l}{\textbf{Case}} & \multicolumn{1}{l}{$\Delta$ (hr)} & \multicolumn{1}{l}{\textbf{Case}} & \multicolumn{1}{l}{$\Delta$ (hr)} \\ \hline
HSHDHF0                                 & 13.30                             & HSHDHF1                              & 16.62\\
HSHDLF0                                 & 14.96                             & HSHDLF1                              & 16.62\\
LSHDHF0                                 & 14.96                             & LSHDHF1                              & 18.28\\
LSHDLF0                                 & 16.62                             & LSHDLF1                              & 19.95\\
HSLDHF0                                 & 23.27                             & HSLDHF1                              & 24.93\\
HSLDLF0                                 & 23.27                             & HSLDLF1                              & 24.93\\
LSLDHF0                                 & 23.27                             & LSLDHF1                              & 24.93\\
LSLDLF0                                 & 26.60                             & LSLDLF1                              & 28.26\\ \hline
\label{table: shock_propagation_duration}
\end{tabular}
\end{table}

\subsubsection{Reverse Shock Formation} \label{sec: reverse_shock}

    \begin{figure*}
           \centering
           \includegraphics[width = \textwidth]{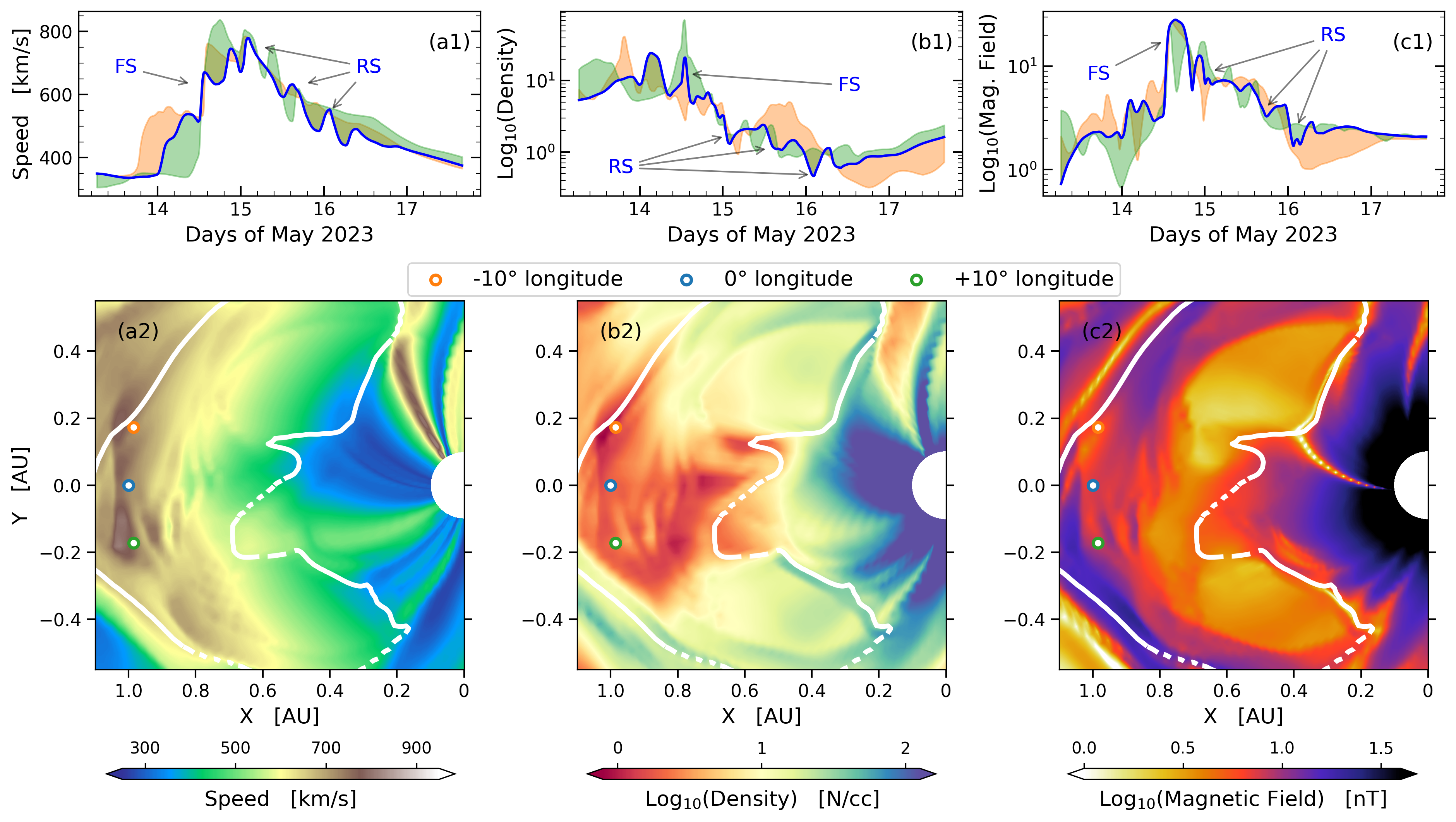}
           \caption{The picture showcases the propagation of reverse shocks inside CME2. The top subplots (a1, b1, and c1) show in-situ measurements from virtual spacecraft as CME2 of the HSHDHF0 case passes through, highlighting forward shocks (FS) and reverse shocks (RS). The bottom subplots display 2D snapshots of (a2) radial velocity (km/s), (b2) density (log scale, N/cc), and (c2) magnetic field strength (log scale, nT). The white line contour indicates the boundary of CME2.}
           \label{fig:reverse_shock}
    \end{figure*}
    
As the trailing shock progresses inside the first magnetic cloud, the rear CME begins colliding with the first. This collision starts in stage 2, where the trailing shock accelerates the rear of CME1. Depending on the advancement of the shock, the difference in speed between the rear of CME1 and the front of CME2 can exceed or fall below the local fast magnetosonic speed. When this speed is exceeded, the trailing CME pushes the plasma faster than the leading CME can smoothly adjust, leading to the formation of a shock wave directed towards CME2.

Fast reverse shocks associated with SIRs and their corresponding in-situ properties have been well studied \citep{kilpua_2015_properties, oliveira_2016_magnetohydrodynamic}. However, their formation due to CME-CME interaction has not been reported until very recently by \cite{trotta_2024_observation} using Solar Orbiter observations. Similar in-situ features have also been observed through virtual spacecraft in our simulation. The bottom panel of Figure \ref{fig:reverse_shock} (a2, b2 \& c2) demonstrates the existence of fast reverse shocks. The characteristic feature of anti-correlation of plasma speed with magnetic field and density can also be seen. As this reverse fast-mode shock wave propagates anti-radially, it compresses the downstream plasma inside CME2, resulting in higher plasma density in the downstream region compared to the upstream region. Conversely, the effective plasma speed will be lower in the downstream region due to the anti-radial direction of the shock, leading to an anti-correlation between speed and density profiles.

 Given the non-uniform interaction demonstrated in earlier sections, the deformation of CME1 causes the collision to initiate in one or more localized areas rather than across the entire interface simultaneously. This non-uniformity leads to the generation of independent shock fronts in each of these areas. The interactions among these shocks can lead to complex dynamics, including the merging of shocks, the amplification or attenuation of wave fronts, and the formation of complex wave patterns. Figure \ref{fig:reverse_shock} illustrates his complexity, showing the reverse shock's propagation in the rear magnetic cloud for the HSHDHF0 case. Inside CME2, a complex pattern of alternating compressed and rarefied zones is visible, manifesting as ripple-like structures in the in-situ plots. Such variations hold the potential to influence the duration of geomagnetic storms.

\subsection{Impact on first CME}

One of the crucial aspects of CME-CME interaction is the alteration in the properties of CMEs, particularly the first CME. This ensemble study, in which the first CME (CME1) remains constant while the second CME (CME2) varies across different cases, offers a unique approach to analyzing how CME1 is influenced by various CME2 scenarios. Drawing from previous studies \citep[][and references therein]{shen_2017_on}, our analysis primarily focuses on changes in total momentum, kinetic energy, magnetic energy, and radial extent of CME1 resulting from its interaction with CME2. The overarching approach is to examine the temporal evolution of impacts on CME1 caused by different CME2 scenarios. To facilitate this, we compared the simulations of a single first CME, case 0, with those from the ensemble cases, cases 1-16. The premise is that any observed changes in CME1 in cases 1-16, relative to the CME in case 0, are due to the influence of CME2.

    \begin{figure*}
           \centering
           \includegraphics[width = \textwidth]{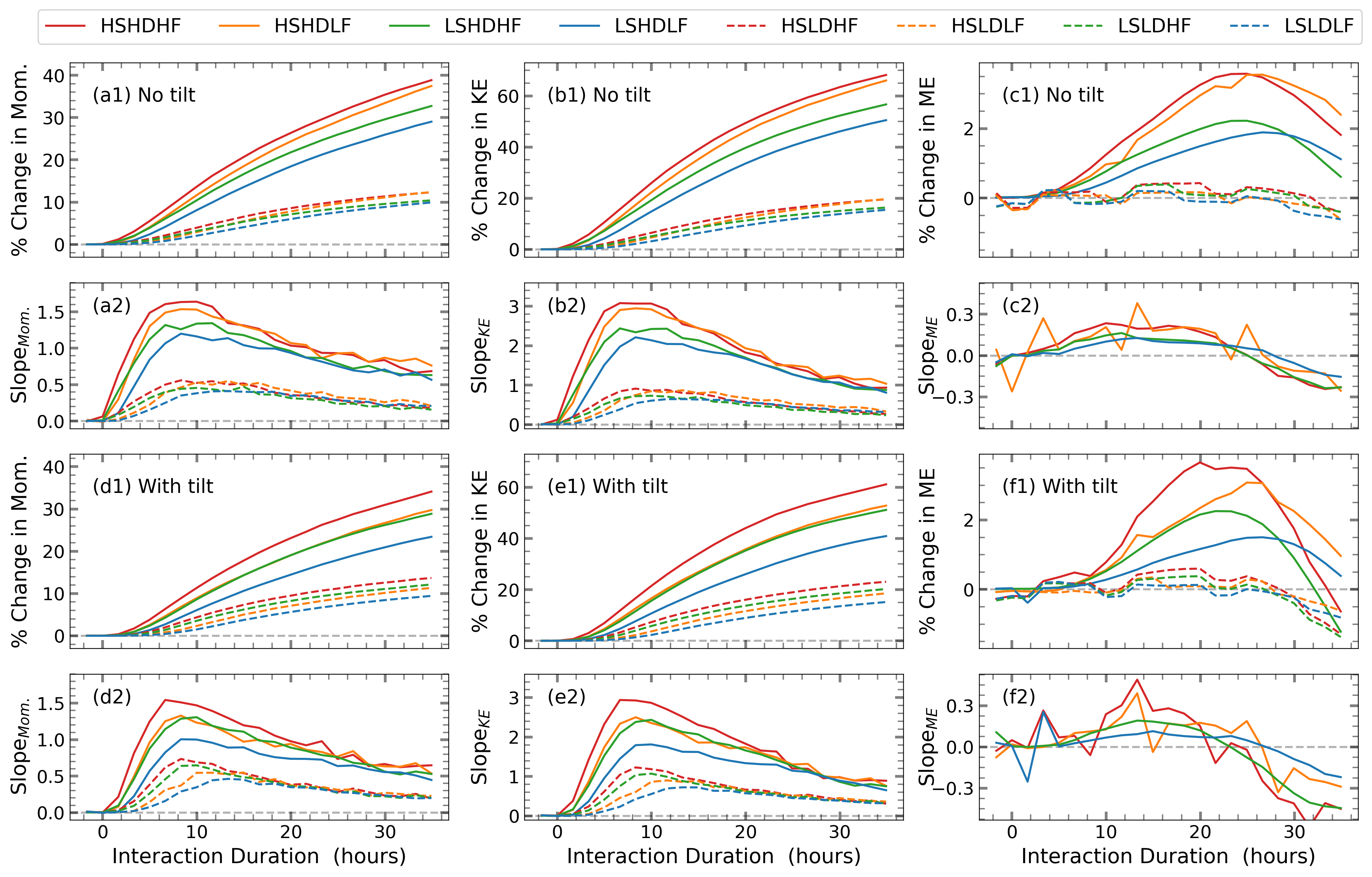}
           \caption{Subplots demonstrates the temporal evolution of change in the radial momentum, total kinetic energy and total magnetic energy of CME1 of cases 1-16, with respect to single CME case.}
           \label{fig:CME1_Impact}
    \end{figure*}

\subsubsection{Kinematics}
The subplots (a1, a2, d1, and d2) of Figure \ref{fig:CME1_Impact} illustrate the percentage change in total radial momentum (TRM) of CME1, which is calculated using following equations:

    \begin{equation}\label{eq:momentum_change}
       \% \, Change = 100 \times \left( \frac{TRM_i - TRM_0}{TRM_0} \right) ,
    \end{equation}
    \begin{equation}\label{eq:momentum}
       TRM = \sum_{j} \rho_j \, v_{r_j} \, dV_j  \,\, .
    \end{equation}

Here, the subscript `i' represents cases 1-16, while `0' denotes case 0. $\rho_j$, $v_{r_j}$, and $dV_j$ are the mass density, radial velocity, and volume of the grid cell `j', respectively. The TRM is the summation of the radial momentum at each grid cell. Based on the slope of the momentum profile, we interpret the temporal evolution of momentum change in CME1 in two distinct phases: \textit{Rising Phase} and \textit{Diminishing Phase}.

The Rising Phase commences when the trailing shock from CME2 impacts the rear of CME1. During this phase, the effective total momentum of CME1 starts to increase due to the local velocity increase of CME1 plasma. As the shock propagates deeper into CME1, the total momentum continues to rise. This momentum increase of CME1 solely due to the trailing shock occurs very briefly (1-2 hours), followed by the collision of the trailing sheath and magnetic cloud with CME1. This collision initiates a substantial momentum exchange between the two CMEs. The high-velocity trailing CME2 exerts significant force on the slower-moving CME1, resulting in a notable increase in CME1's total momentum. During this phase, the rate of momentum increase (indicated by the slope of the momentum curve) achieve its highest value, reflecting the extent of direct and robust transfer of momentum from CME2 to CME1.

In the final Diminishing Phase, CME2 has transferred a considerable amount of its momentum to CME1, significantly reducing its own momentum. This momentum exchange decreases the relative speed between the two CMEs. Despite the reduction, CME2 still maintains a higher speed than CME1 and continues to push it forward, albeit with diminishing force. Consequently, the rate of change in momentum for CME1 gradually decreases, indicating that the interaction force is waning and the system is approaching a new equilibrium state.

The transition between phases is quite smooth, making it challenging to define rigid boundaries. Relatively speaking, the Rising Phase has an average duration of approximately 10 hours, after which the Diminishing Phase persists. The exact duration of this transition varies across different cases. For the HSHDHF case, the Diminishing Phase starts earliest, at roughly 7 hours, whereas for the LSLDLF case, it takes the longest time, approximately 13 hours. For cases with relative tilt between CMEs, this transition period is slightly shorter compared to cases with no relative tilt.

The momentum gained by CME1, from the start of the interaction to the phase transition point and until the combined structure reaches 1 AU, also varies significantly with the initial conditions of CME2. The 8 cases shown in the subplots can be clustered into two groups: high-density (HD) and low-density (LD) cases. After 10 hours of interaction between CME1 and CME2, CME1 in HD cases gains approximately 5-15\% more momentum, whereas in LD cases, it gains 1-5\% more momentum. This momentum gain is slightly higher for cases with no relative tilt compared to cases with relative tilt. Moreover, the difference between the HD and LD cases is also more pronounced for no relative tilt cases.

As the CME-CME structure reaches 1 AU, starting from 25 hours, the difference in momentum gain between HD and LD cases becomes more pronounced. Since shock strength is inversely correlated with shock propagation time, which is shorter for high-density cases (see Table \ref{table: shock_propagation_duration}), these cases exhibit higher momentum gain. Additionally, CME2 in HD cases has more initial momentum to impart than in LD cases. After 30 hours of interaction, CME1 in HD cases demonstrates a 20-40\% increase in momentum compared to 7-12\% in LD scenarios. The gap between these two clusters of cases becomes 16\% in the absence of relative tilt compared to 8\% in the presence of tilt.

Since the total kinetic energy (KE = 0.5$\sum_{j} \rho_j , v_{r_j}^2 , dV_j$) of a radially evolving CME1 is related to its total radial momentum, a similarity between the two is expected. The subplots (b1, b2, e1, and e2) of Figure \ref{fig:CME1_Impact} depict the temporal evolution of the percentage change in the KE of CME1 for cases 1-16 relative to case 0. These subplots closely resemble the momentum plots but indicate a greater gain. On average, the percentage gain in KE is about two-thirds greater than the percentage gain in momentum of CME1 due to the interaction process. As with momentum, the eight cases can be clustered into two groups based on their initial density. The difference between these two groups is more pronounced in the absence of relative tilt (up to 25\%) compared to scenarios with non-zero tilt (up to 18\%). After 30 hours of interaction, CME1 gains up to 60\% more kinetic energy in the HSHDHF case and up to 20\% in the LSLDLF case due to the interaction with CME2.

\subsubsection{Magnetic Energy}  \label{sec:magnetic energy}
The evolution of total magnetic energy (ME) of CME1 differs significantly from the changes observed in momentum and kinetic energy. The subplots (c1, c2, f1, and f2) of Figure \ref{fig:CME1_Impact} showcases the temporal evolution of the percentage change in ME of CME1. These subplots reveal distinct behaviors based on the initial density of CME2. In low-density cases, the interaction has little to no effect on the ME of CME1. In contrast, high-density cases exhibit an initial increase in ME of approximately 1-3\%, followed by a subsequent decrease. This pattern is consistent in both tilt and no-tilt scenarios, although the changes in ME occur more rapidly in the presence of relative tilt.

As the interaction begins, CME2 compresses CME1, potentially increasing the magnetic field strength and, consequently, the total magnetic energy if magnetic flux is conserved. Since CME1 and CME2 have the same chirality, this compression could also induce magnetic reconnection and other instabilities (e.g., tearing mode and plasmoid instabilities), dissipating magnetic fields and converting magnetic energy into kinetic and thermal energy. Despite significant compression even in the LSLDLF case, the gain in ME for low-density scenarios is almost negligible, suggesting the dissipation of magnetic field may happen even in weaker collisions.

In high-density cases, there is a noticeable increase in ME, implying that the rate of magnetic field strength enhancement due to compression initially exceeds the rate of magnetic dissipation. However, after approximately 25 hours of interaction, the ME begins to decrease, indicating that the rate of compression diminishes while the rate of magnetic dissipation becomes dominant, leading to a reduction in the total magnetic energy of CME1.

Unlike momentum and kinetic energy profiles, ME profiles can be categorized into three groups: low-density (LD) cases, high-density high-flux (HDHF) cases, and high-density low-flux (HDLF) cases. Given the insignificant changes in ME for LD cases, their slopes are excluded from subplots (c2 and f2) to highlight the behavior of the remaining cases. Comparing HSHDHF (red) and HSHDLF (orange) cases, the initial rise in ME is greater for the HSHDHF case, but the subsequent decrease is also more rapid. After 30 hours of interaction, the effective ME is lower for the HSHDHF case. This trend is consistent when comparing LSHDHF to LSHDLF cases and across tilt and no-tilt scenarios. This suggests that a higher initial magnetic flux results in a greater initial gain in ME due to compression, followed by a more substantial decrease in ME due to magnetic dissipation.

\subsubsection{Radial Extent}

In the earlier sections, we demonstrated and discussed the non-uniform radial expansion of CME1 due to interactions with the solar wind from the front and CME2 from the rear (see Figures \ref{fig:2D_example} and \ref{fig:2D_shock}). We also noted in the last section that the only possible way of increasing the magnetic energy is the compression of CME1 due to CME2. Although multiple studies \citep{xiong_2006_magnetohydrodynamic, lugaz_2013_the, lugaz_2017_the} have examined the compression of the leading CME in CME-CME interactions, a quantitative analysis of the temporal evolution of this compression has not been performed.

    \begin{figure}
           \centering
           \includegraphics[width = \columnwidth]{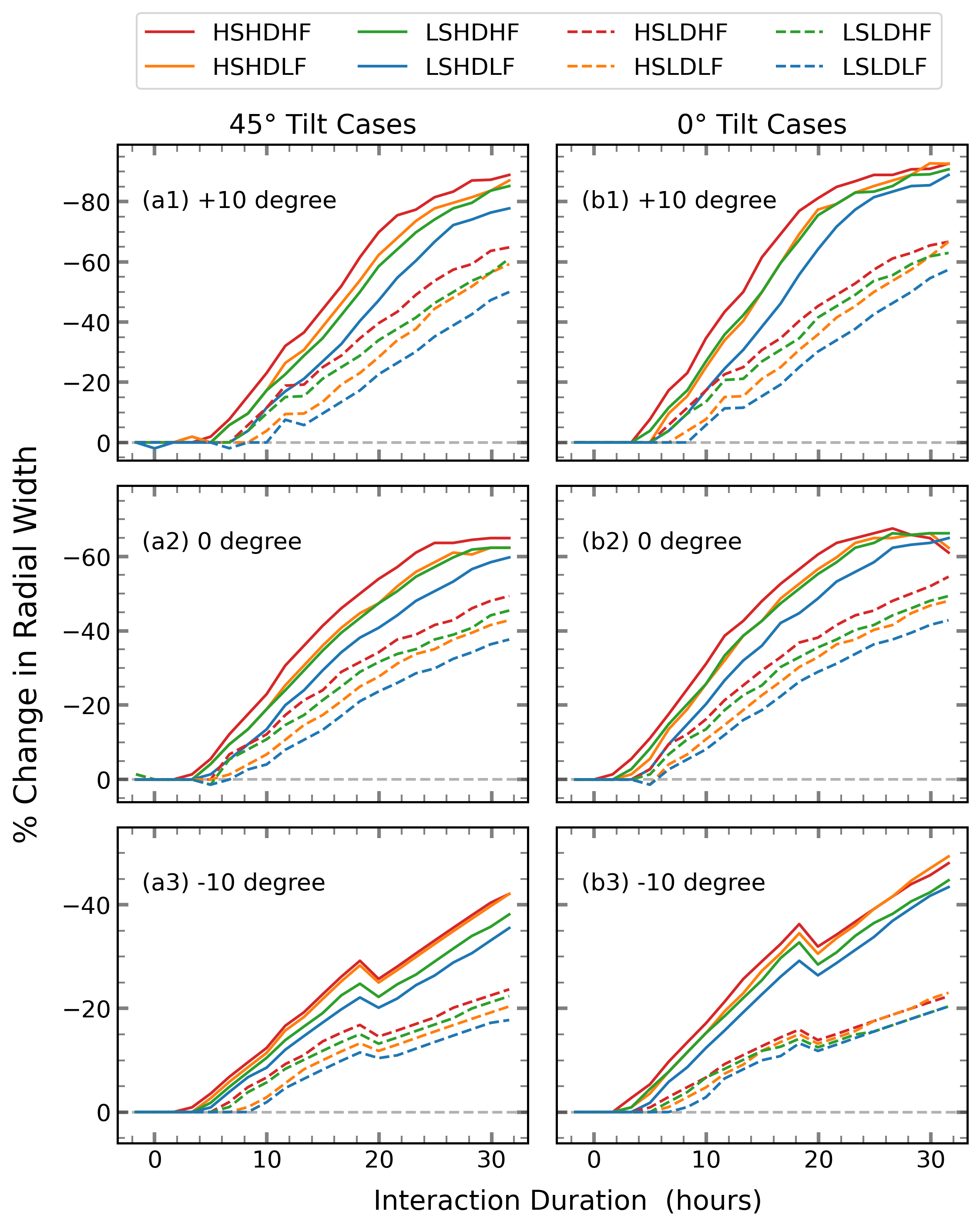}
           \caption{Temporal evolution of the percentage change in the radial width of CME1 compared to the single CME case. Results are presented along three longitudes (0\textdegree\, and $\pm$10\textdegree) for all cases.}
           \label{fig:CME1_radial}
    \end{figure}

Using the passive scalar tracing technique, we traced the CME1 structure and computed its radial compression along different longitudes. This method tracks plasma using the advection equation, allowing the study of transport and mixing without altering magnetic or fluid dynamics, as demonstrated in \citep{prateekmayank_2023_swasticme}. By placing virtual spacecraft at the front and rear of CME1, we could measure the radial difference between them, thus determining the radial extent of CME1 along each longitude. 
Figure \ref{fig:CME1_radial} shows the temporal evolution of the percentage change in CME1's radial width across all CME-CME interaction cases (1-16) versus the single CME case, with maximum compression seen in the HSHDHF case and minimum in the LSLDLF case across -10\textdegree, 0\textdegree, and +10\textdegree.

Similar to the momentum, kinetic energy, and magnetic energy profiles, the compression features of HD and LD cases are distinguishable in subplots of Figure \ref{fig:CME1_radial}. Compression is consistently higher for HD cases from the onset, though no noticeable gap between LSHDLF and HSLDHF cases appears until about 10 hours for the 45\textdegree\ tilt cases. At +10\textdegree, (\ref{fig:CME1_radial}a1) HD cases with 45\textdegree\ tilt (HD1) reach an 80\% reduction over 30 hours, while LD1 shows 60\% reduction. Along -10\textdegree, (\ref{fig:CME1_radial}a2) HD1 shows a 40\% reduction and LD1 a 20\% reduction. he absence of an SIR along this longitude means there is no significant obstruction to CME1's expansion from the front, resulting in less compression from the rear than at +10\textdegree, with a nearly constant decrease for all cases. At 0\textdegree, (\ref{fig:CME1_radial}a3) HD1 cases shows a 65\% reduction compared to 45\% for LD1, revealing that the presence of the SIR ahead of CME1 has intermediate ipact of this region, more than -10\textdegree\ but less than +10\textdegree.

The above-mentioned statistics pertain to cases with a 45\textdegree\ tilt between the CMEs. In cases with a 0\textdegree\, tilt (see Figure \ref{fig:CME1_radial}, panels b1-b3), the interaction begins slightly earlier, leading to faster and greater compression. On average, the compression in cases without tilt is approximately 5\% greater than in cases with tilt. This suggests that the alignment of CME2 with CME1 can influence the efficiency of the compressive interaction. Due to this tilt-induced difference, an enhancement in the gap between HD and LD cases is evident in Figure \ref{fig:CME1_radial}. When comparing the overall effect of CME2's initial properties, the HSHDHF case consistently shows the greatest compression, followed by HSHDLF and LSHDHF cases.
    
\subsection{Mixing of CMEs} \label{sec: mixing}

In addition to the merging of shocks and alterations in CME properties, interactions between magnetic clouds can also lead to their merging. Multiple observational studies have suggested \citep{gopalswamy_2001_radio, burlaga_2002_successive} and demonstrated \citep{lugaz_2009_deriving, temmer_2012_characteristics, liu_2012_interactions} the merging of magnetic clouds in the inner heliosphere during CME-CME interactions. However, the extent of plasma mixing between two interacting CMEs remains largely unknown. Particularly, a quantified study of this mixing and its dependence on other CME properties has not been conducted yet. Understanding this mixing is crucial, as it may influence the geoeffectiveness of CMEs, impacting the intensity and duration of geomagnetic storms.

\subsubsection{Quantifying the Mixing of CMEs}

To estimate the extent of plasma mixing between the leading and trailing CMEs, we utilized the CME tracer described in earlier sections to define the boundary of the CME structure. This approach is analogous to methods used in studying the mixing of astrophysical jets \citep[e.g., see][]{walg_2013_relativistic}. To quantify this mixing, we define a \textit{mixing factor} ($\mathcal{M}$), which represents the absolute mass fractions within a specific grid cell. We set $\mathcal{M} = 0$ for the case of no mixing, where only CME1 material is present, and $\mathcal{M} = 2$ for the case where only CME2 material is present. We set $\mathcal{M} = 1$ in the case of maximum absolute mixing, meaning equal amounts of CME1 and CME2 constituents are present within the grid cell. In this scenario, the mass fraction of CME1 is equal to the mass fraction of CME2.

At a given time ($t$) and distance (\textbf{r}), a tracer $\mathcal{T}(t, \textbf{r})$ is advected by the flow and obtains values within the range $\mathcal{T}_{\min} \leq \mathcal{T}(t, \textbf{r}) \leq \mathcal{T}_{\max}$. Here, $\mathcal{T}_{\min}$ corresponds to the absence of that CME within the cell, while $\mathcal{T}_{\max}$ corresponds to a cell purely containing the plasma of that CME. In this work, we have taken $\mathcal{T}_{\min}$ = 0.1 and $\mathcal{T}_{\max}$ = 1.0 for all CMEs.
Given that a tracer value directly corresponds to the quantity of that CME, the mass fraction of CME1 ($\delta_1$) within a grid cell can be expressed as follows:

    \begin{equation}\label{eq:mass_frac}
       \delta_1(t, r) = \left| \frac{\mathcal{T}_1(t, r) - \mathcal{T}_{\min}}{\mathcal{T}_{\max} - \mathcal{T}_{\min}} \right|  .
    \end{equation}

Similarly, the mass fraction of CME2 ($\delta_2$) can also be calculated. Furthermore, assuming that the mass fraction within a cell linearly scales with the amount of mixing, the mixing factor of the plasma of CME1 and CME2 in terms of their tracer values in a grid cell can be written as:

\begin{equation}\label{eq:mixing_factor}
       \mathcal{M} = 1 - \left( \frac{\mathcal{T}_1 - \mathcal{T}_2}{\mathcal{T}_1 + \mathcal{T}_2} \right)  .
    \end{equation}

\subsubsection{Analysing the Mixing of CMEs}

\begin{figure}
           \centering
           \includegraphics[width = \columnwidth]{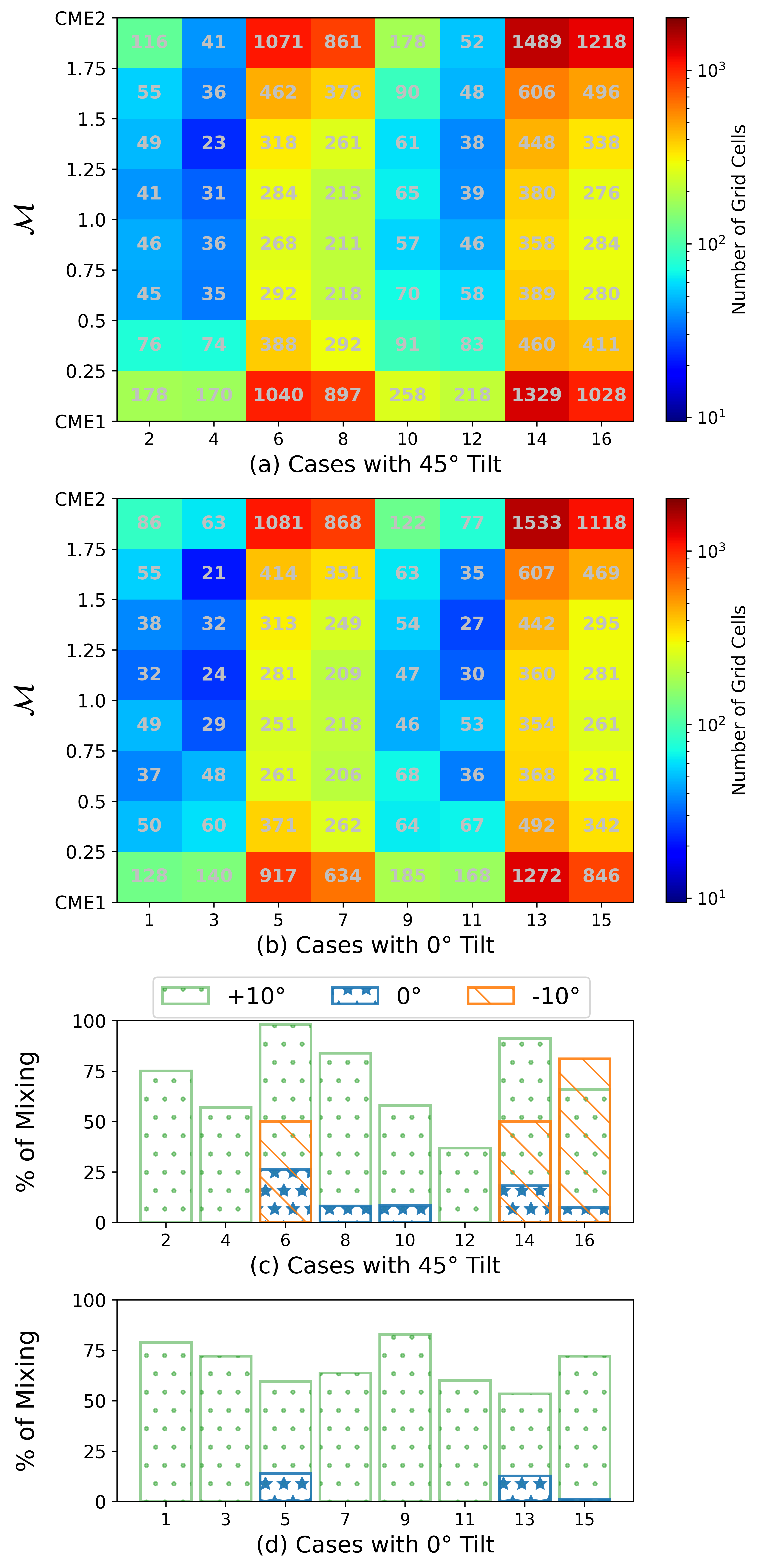}
           \caption{Amount of mixing between CME1 and CME2 upon the interacting structure's arrival at 1 AU. Subplots (a) and (b) show the mixing factor ($\mathcal{M}$) for cases with 45\textdegree\ and 0\textdegree\ relative tilt between the CMEs, along with the number of grid cells corresponding to each bin. Subplots (c) and (d) present the mean percentage of CME1-CME2 mixing along the three longitudes (0\textdegree\, and $\pm$10\textdegree) in the equatorial plane.}
           \label{fig:mixing}
    \end{figure}

Figure \ref{fig:mixing} presents the 2D histogram of the calculated mixing factor values throughout the entire simulation domain for all ensemble cases at the time when the merged structure reaches 1 AU. These values are computed for each grid cell within the computational domain and binned into eight uniform clusters, ranging from 0 to 2. A mixing factor of $\mathcal{M} = 1$ indicates the highest level of mixing (50\%), meaning both CME1 and CME2 contribute equal amounts of plasma to that specific cell. The cases with and without relative tilt between the leading and trailing CMEs are plotted separately. The color of each block represents the number of grid cells associated with a specific amount of mixing (Y-axis) for each corresponding case (X-axis). For instance, in case 15, 261 grid cells have $\mathcal{M}$ values between 0.75 and 1 (see Figure \ref{fig:mixing}(b)).

The most notable feature in these subplots corresponds to the high-density cases (5 to 8 and 13 to 16), which exhibit a significantly larger volume of mixing. As discussed earlier, these scenarios also demonstrated stronger CME shocks, higher momentum exchange, and greater radial compression. This trend is consistent in terms of the extent of mixing, where $\mathcal{M}$ values are several times higher in high-density cases as compared to low-density cases. High-speed cases (9 to 16) also consistently exhibit greater mixing compared to low-speed cases. Although the difference is not as big as in high versus low-density cases, the trend is evident across all bins of mixing factor.

Apart from these global trends, we also observed non-uniform mixing across the CME2 front. Similar to the non-uniform interaction between the back of CME1 and the front of CME2 discussed in earlier sections, the bottom flank exhibited higher percentage of mixing (100 $\times$ [1 - $|\mathcal{M}-1|$]) compared to the upper flank. Panels (c) and (d) of Figure \ref{fig:mixing} show the mean percentage of mixing  along +10\textdegree\ (bottom flank), -10\textdegree\ (upper flank) and 0\textdegree\ longitudes for 45\textdegree\ and 0\textdegree\ cases, respectively. In these plots, the +10\textdegree\ region consistently exhibits non-zero mixing, with highest percentage in almost all the cases, suggesting that the bottom flank of CME2 experiences more intense interaction with CME1. In contrast, along -10\textdegree, only three cases with a 45\textdegree\ tilt and none with a 0\textdegree\ tilt show any mixing, while the 0\textdegree\ region exhibits mixing in a total of eight cases. This pattern supports the observed asymmetry in interaction strength, where the bottom flank experiences stronger compression, enhancing the mixing process. This again highlights the impact of the inhomogeneous ambient solar wind, which causes non-uniform interaction and mixing of the interacting CMEs.

\section{Geo-effectiveness} \label{sec:geo-effectiveness}

    \begin{figure*}
       \centering
       \includegraphics[width = \textwidth]{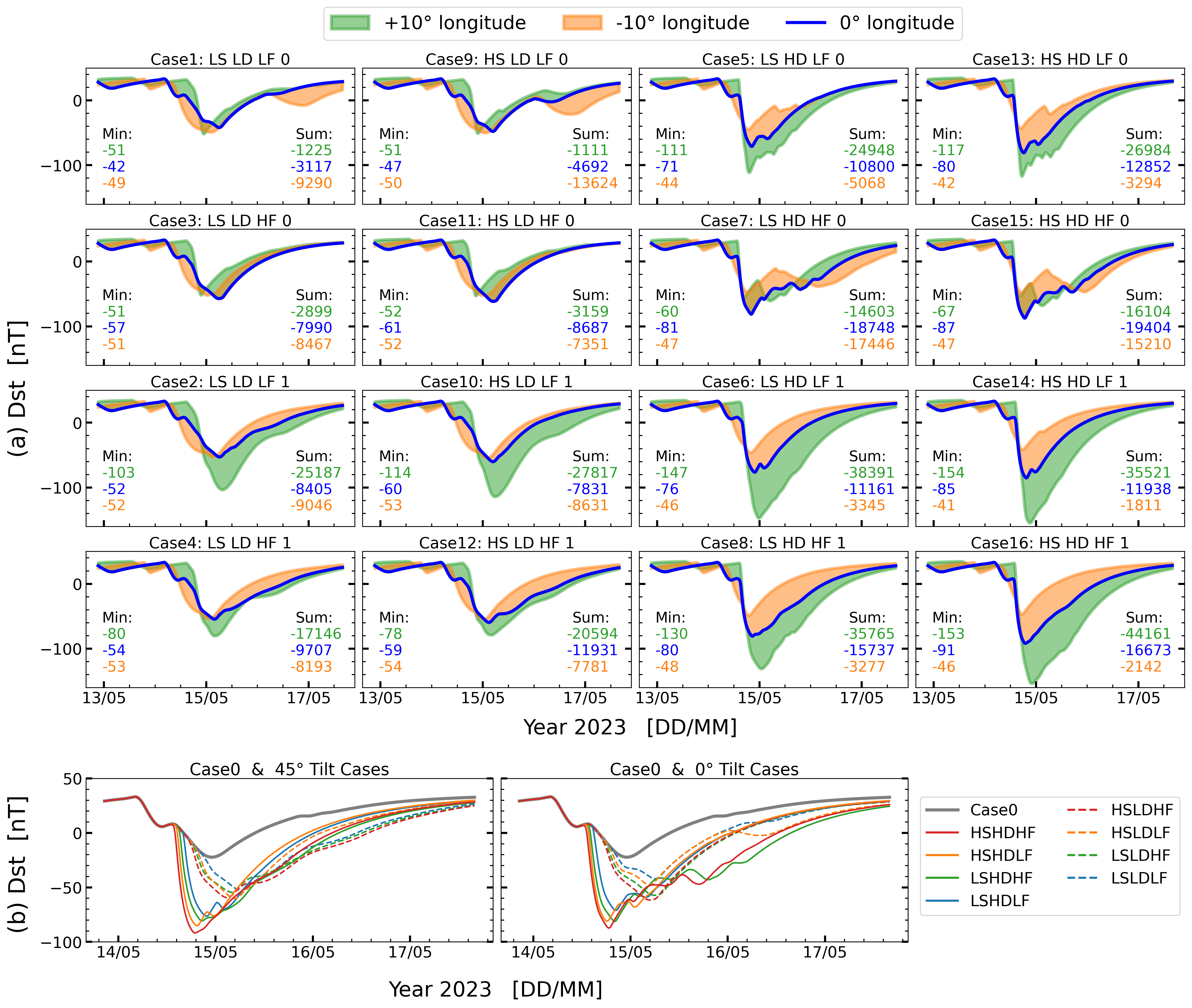}
       \caption{The Dst indices of all simulated cases in the ensemble, case 1-16, at three longitudes at 1AU. Each panel in (a) shows the time series of the Dst index for different cases. The values of ``Min" indicate the minimum Dst value during the event for each longitude, while ``Sum" represents the integrated Dst index over time. Subplots (b) at the bottom compare the Dst indices at 0\textdegree\ longitude of Case 0 (single CME case) with the cases having 45\textdegree\ and 0\textdegree\ tilt between the CMEs.}
       \label{fig:Dst_all}
    \end{figure*}

In the previous section, we discussed the evolution of CME-CME interactions and their impact on the kinematic, magnetic, and structural properties of CME1. From a geo-effectiveness perspective, it is crucial to understand how this evolution translates into in-situ properties at 1 AU. Specifically, we aim to determine how the interactions between two CMEs ultimately transform into plasma properties that will interact with the Earth's magnetosphere. To investigate this in depth, we placed three virtual spacecraft in our simulation at 0\textdegree\ and $\pm$10\textdegree\ longitudes, as depicted in Figures \ref{fig:2D_example} and \ref{fig:2D_shock}.

The simulated time-series data, including speed, density, and Bz profiles at a 5-minute cadence, extracted from the virtual spacecraft positions, are presented in the Appendix. The features observed in the evolution of CME1 and CME2 are reflected in their in-situ properties, which in turn lead to variations in the Dst profile. The next subsections delve into these changes in Dst, focusing specifically on variations in the overall trend, the minimum Dst value, and the cumulative Dst.

\subsection{Dst Variation}\label{sec: dst_variation}

Figure \ref{fig:Dst_all}(a) showcases the variations in Dst values across the 16 cases at three different longitudes: +10\textdegree \, (green), 0\textdegree \, (blue), and -10\textdegree \, (orange). Shaded regions depict the difference between the values at $\pm 10^\circ$ (green \& orange) and $0^\circ$ (blue). The \ref{fig:Dst_all}(b) subplots present the temporal overlap of the Dst profiles corresponding to the 16 CME-CME interaction cases, along with the single CME case (Case 0). The plotted Dst values are computed based on the method described in Section \ref{sec: Dst_model}. The analysis of these Dst profiles can be segmented into two distinct phases: the main phase and the recovery phase, which is divided by the global minimum in the Dst profile. The effects of the varying strengths of shocks, sheath regions, and CME1 are primarily manifested in the main phase. On the other hand, differences in the nature of the trailing CME are reflected in the recovery phase. The following discussion explores these significant impacts and their connections to the in-situ signatures.

\subsubsection{Main Phase}

The onset of the Dst main phase begins with the arrival of the first shock at 1 AU. Figure \ref{fig:Dst_all} illustrates the complex variations among the 16 cases, highlighting the different starting and ending times of the main phase. The onset time variation primarily depends on the ambient solar wind conditions, which remains constant across all cases, resulting in similar onset times for different cases at the same longitude. However, in scenarios where shock-shock interactions occur, deviations from the initial onset times are observed. Since the shock first impacts the spacecraft at +10\textdegree\ longitude and no shock-shock mergers occur at +10\textdegree\ in any of the cases, the main phase starting time is identical for the green profile across all cases. This timing aligns with the arrival of the first shock ($\sim$20:00 hours on 13th May 2023).

Similar to the +10\textdegree\ longitude, the two shocks arrive sequentially at 0\textdegree\ longitude at 1 AU without merging in any of the cases. Both locations exhibit a two-dip profile, where the first dip is associated with the arrival of the first shock and the second dip begins with the arrival of the second shock. In the orange profile, the first Dst drop is very small (around 20:00 UT on 13/05) and recovers almost fully to the pre-storm level within a few hours, just before the second shock arrives at about 12:00 UT on 14/05. Depending on the strength of the second shock, the Dst value then drops again, reaching a minimum either more rapidly or gradually.  
A similar trend is observed for the blue profile, with the main difference being that it takes longer for Dst to recover to pre-storm levels before the arrival of the second shock.
This variation is primarily due to the B$_{\rm z}$ profiles at these two locations, as shown in Figure \ref{fig:Bz_all}. 
In the orange profile, the negative B$_{\rm z}$ quickly ($\sim$ 4 hours) turns positive, whereas at 0\textdegree\ longitude, it remains negative for almost 7 hours.

Unlike the orange and blue profiles, the Dst main phase profiles at +10\textdegree\ show only a single dip in all cases. In low-density (LD) cases, this dip features two distinct slopes: an initial gradual decline followed by a sharper fall. For high-density (HD) cases, there is a single sharp decline to the global minimum, representing the shock-shock merger and indicating that the CME-CME interaction at this longitude is in its fourth stage. Despite the sharpest decline occurring at this location, the minimum Dst values do not always correspond to this longitude. Interestingly, this is especially not true for high-flux (HF) cases without tilt, where the initial magnetic flux of CME2 is greater. In these cases, the minimum Dst value is seen at 0\textdegree\ (blue line) rather than at +10\textdegree\ (green line). Conversely, in low-flux (LF) cases without tilt, the minimum Dst value appears in the green +10\textdegree) profile. This trend is intriguing because the interaction is most prominent along the +10\textdegree\ longitude. Yet, an increase in the initial magnetic flux of CME2 leads to higher Dst values along this longitude.

There are also some peculiar trends in the main phase corresponding to the initial properties of CME2. Panel \ref{fig:Dst_all}(b) demonstrates clear differences between the HD and LD cases, showing a trend similar to that observed in Figure \ref{fig:CME1_Impact}. The divergence between these cases begins around 15:00 hrs on 14th May with the arrival of the trailing shock, imitating the differences shown in the strength of shocks in Figure \ref{fig:shock_duration}(b). Notably, by the time Dst reaches its global minimum, the temporal gap between HD and LD cases further widens, with HD cases reaching their minimum Dst value before the single CME case.

Additionally, cases with and without relative tilt exhibit significant differences. In most cases, both the cumulative Dst and minimum Dst values are higher when there is a relative tilt between the CMEs compared to when there is none. The primary reason for this is the change in orientation of the magnetic field of CME2, which should ideally enhance the effect along -10\textdegree\ and decrease it along +10\textdegree, while remaining the same at 0\textdegree. However, due to the presence of solar wind and prolonged interaction with CME1, this trend deviates slightly from the ideal scenario. Figure \ref{fig:Bz_all} shows the B$_{\rm z}$ values for all cases, where the minimum B$_{\rm z}$ values exhibit a mixed trend along -10\textdegree, 0\textdegree, and +10\textdegree\ longitudes with changes in tilt. However, the cumulative duration of negative B$_{\rm z}$ is consistently longer for +10\textdegree. Additionally, another noticeable difference is in the slope of the Dst fall, which is steeper for HD cases compared to LD cases.

\subsubsection{Recovery Phase}

In an ideal single CME storm, the recovery phase is smooth and continuous, with the Dst index value generally increasing following an exponential trend and gradually returning to the pre-storm level (as in equation \ref{eq:Dst3}).
Any deviations or additional fluctuations from this ideal trend can indicate the influence of second CME and subsequent solar wind. Figure \ref{fig:Dst_all}(b) showcases that among the 16 ensemble cases, some exhibited significant deviations, others showed minor variations, and a few had changes that were almost negligible.

The high-density (HD) cases exhibit significant deviations from the idealized scenario of the Dst recovery phase. Notably, scenarios without relative tilt between CME1 and CME2 emphasize these deviations more clearly (see Figure \ref{fig:Dst_all}). In cases 5, 7, 13, and 15, a discontinuity is observed in the Dst index increase across all three profiles (orange, blue, and green), resulting in the formation of a recovery phase plateau. 
These interruptions, while not affecting the minimum Dst value, significantly prolong the overall recovery time, delaying the return to pre-storm Dst levels. 
The mentioned HD cases demonstrate notably longer recovery phases compared to their low-density (LD) counterparts. 
For instance, by May 16th, the Dst index had almost recovered to 0 nT for cases 3 and 11 (LDHF0), whereas for cases 7 and 15 (HDHF0), the Dst index remained near -50 nT.

The recovery phase plateau can be attributed to fluctuations in the southward magnetic field as southward IMF excursions drive the Dst down before recovery can resume.
In the B$_{\rm z}$ profile (see Figure \ref{fig:Bz_all}), such fluctuations are noticeable after approximately 18:00 hours on May 14th for HD cases without tilt. These fluctuations in the B$_{\rm z}$ profile are mirrored by similar, more pronounced fluctuations in the in-situ speed profile. As discussed in Section \ref{sec: reverse_shock}, these fluctuations result from the reverse shock formed by the collision of CME2 with CME1. The propagation of this reverse shock through CME2 creates complex patterns of alternating compressed and rarefied regions, leading to the ripples observed in the in-situ profiles. Consequently, these ripples extend the Dst recovery phase by forming a plateau, significantly delaying the return to pre-storm levels.

Apart from the major deviations from the idealized scenarios, many cases showcased minor or no deviations. For example, the blue profile in cases with tilt shows slight deviations; in cases 6 and 14, there is a break point, and in other cases, the profile does not increase smoothly but rather with some disruptions. Notably, there is no significant flipping in the B$_{\rm z}$  profile after reaching the global minimum in the Dst profile around May 15th for these cases. Interestingly, although there are a few break points around May 15th in the speed profile, the absence of flipping in the B$_{\rm z}$  profile prevents the formation of any plateau. In other cases, such as 3 and 11, the blue profile closely follows the idealized scenario. Similarly, the orange profile exhibits an ideal recovery phase in almost all tilt cases. In these scenarios, there is neither directional flipping in the B$_{\rm z}$  profile nor any break points in the speed profile, resulting in a smooth, continuous function.

\begin{figure*}
       \centering
       \includegraphics[width = \columnwidth]{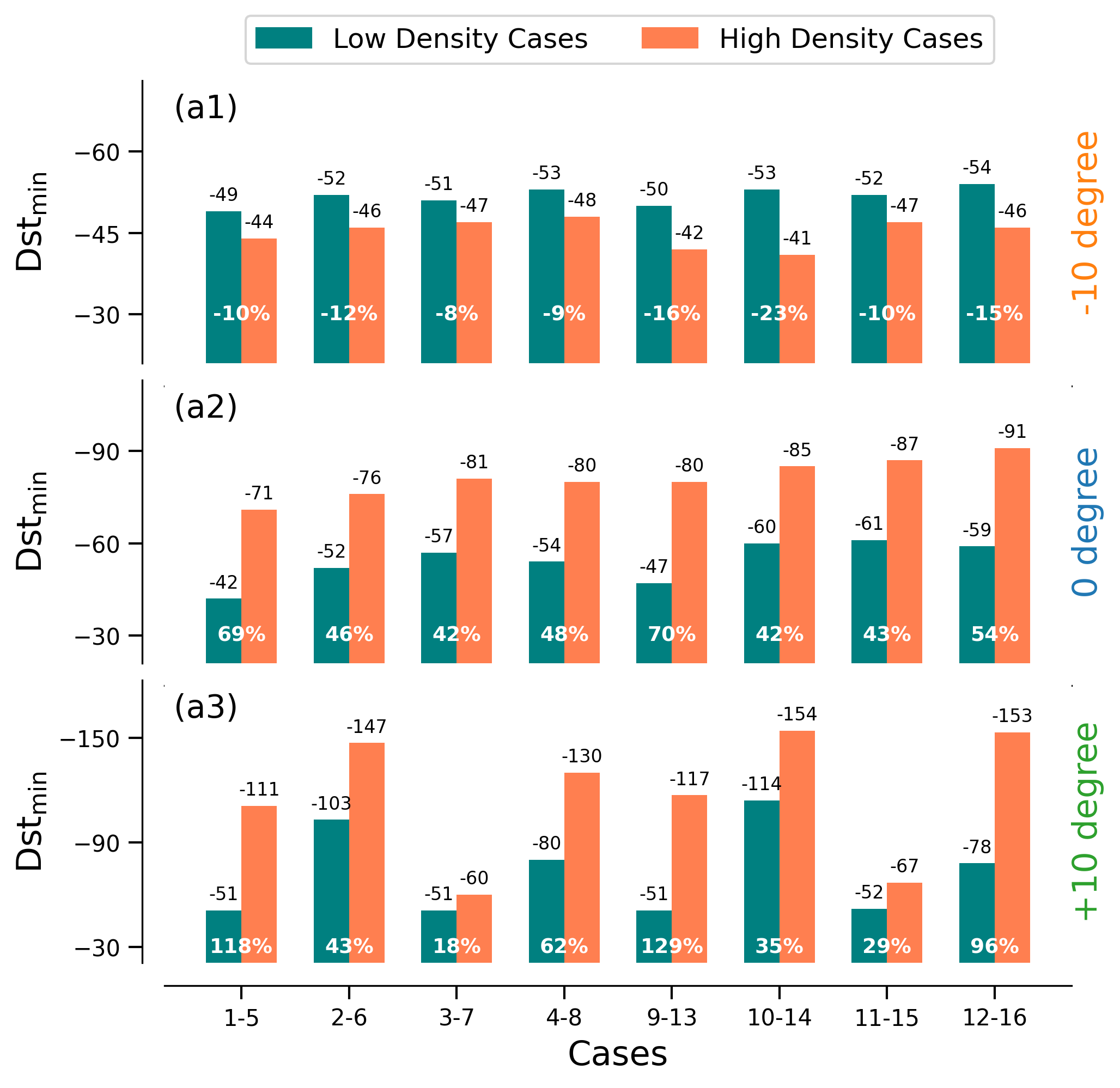}
       \includegraphics[width = \columnwidth]{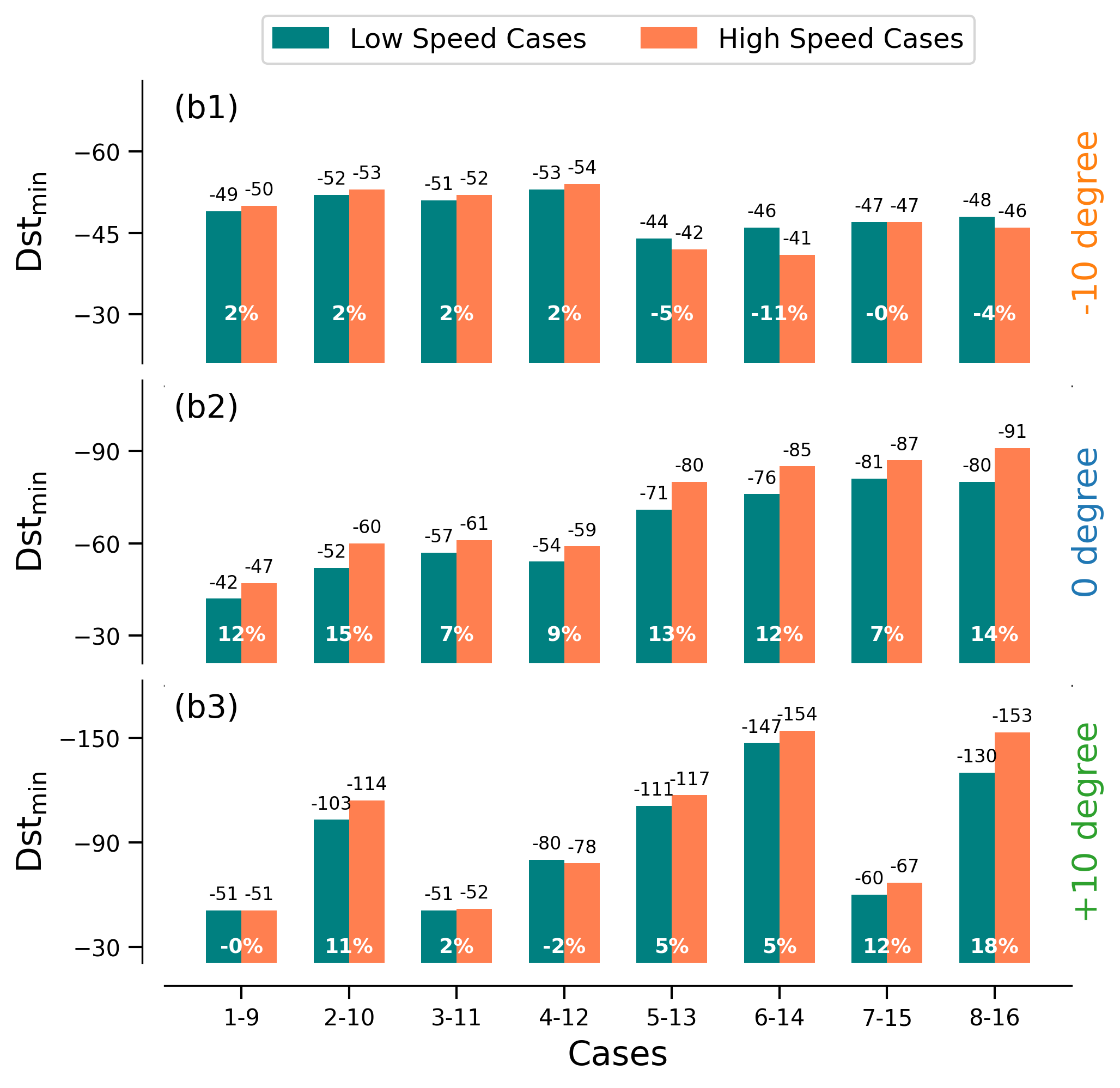}
       \includegraphics[width = \columnwidth]{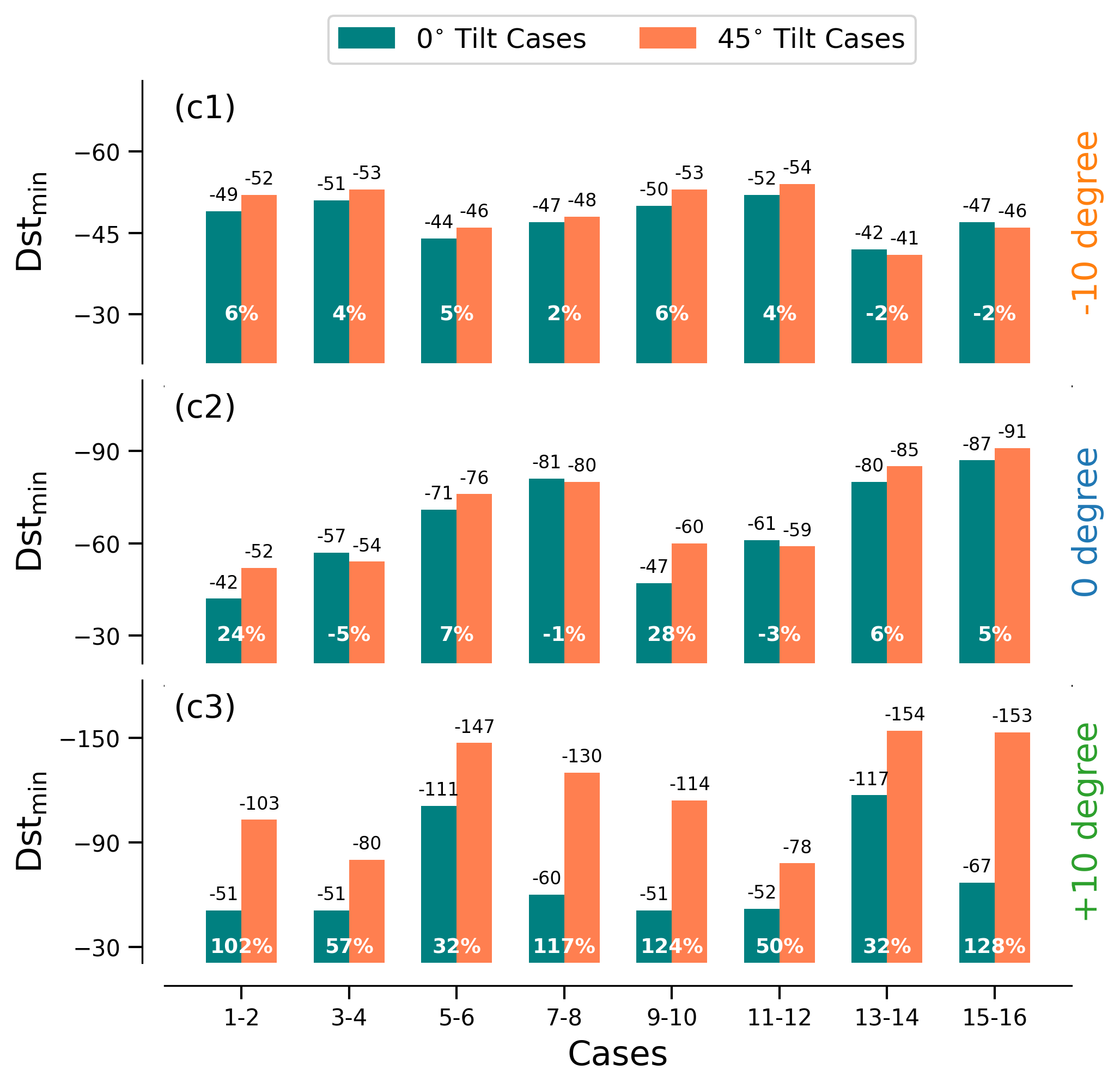}
       \includegraphics[width = \columnwidth]{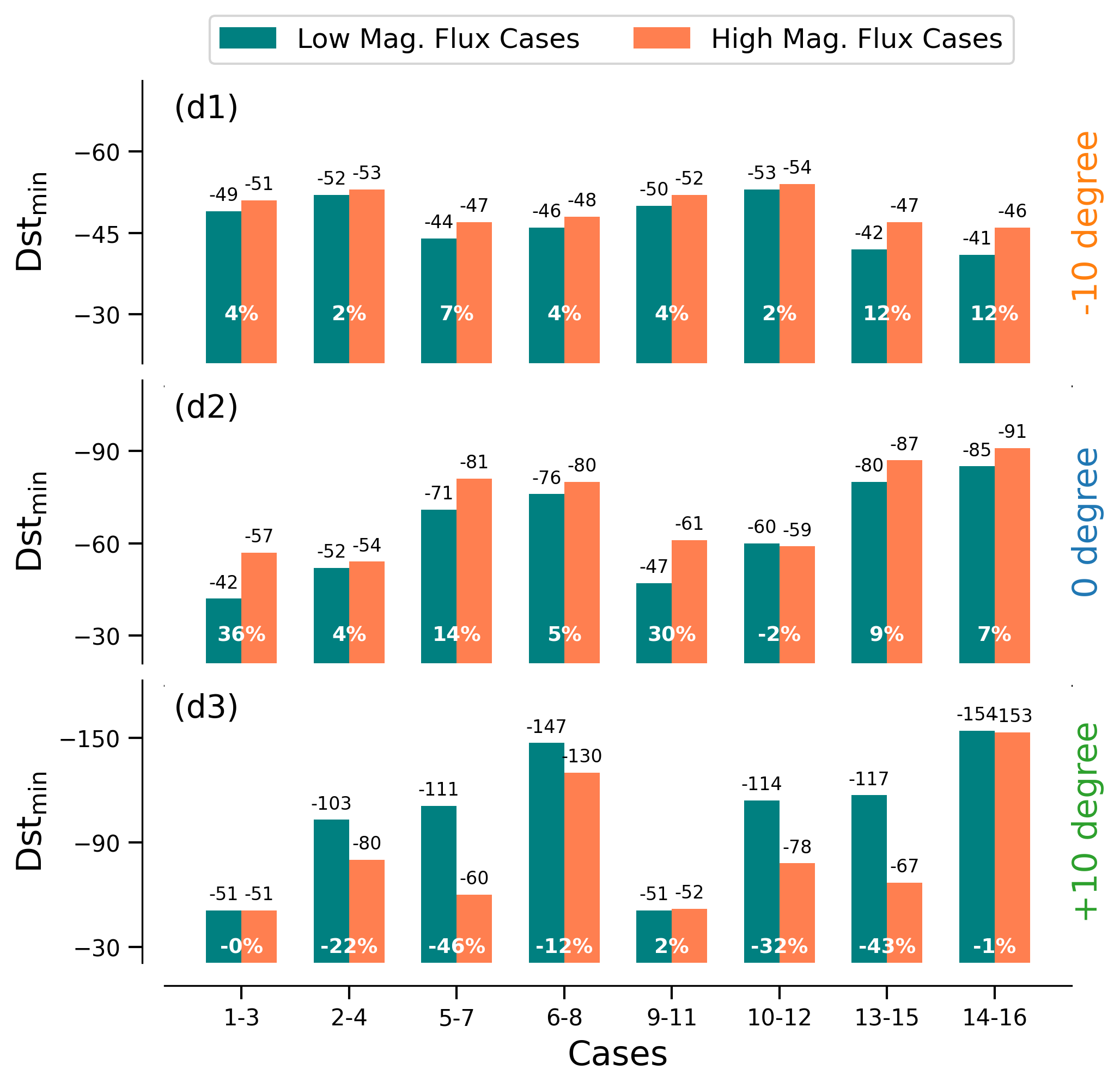}
       \caption{The histogram plots of the minimum Dst value corresponding to the change in initial density (a1-a3), speed (b1-b3), tilt (c1-c3) and magnetic flux (d1-d3), at three longitudes at 1AU.}
       \label{fig:Dst_min}
\end{figure*}

\subsection{Minimum Dst Index} \label{sec: Dst_min}

Dst is a measure of the storm time ring current and the minimum value of Dst reached during a storm is routinely used as a proxy for storm intensity \citep[][and references therein]{Borovsky2017}.
Although it is by no means a complete characterisation of storm time activity we use this index as a quantifier of storm intensity -- specifically the minimum value of estimated Dst indices at 1AU.
To identify trends accurately, we have grouped the cases into pairs, where each pair differs by only one parameter: initial density, speed, flux, or tilt of CME2. This approach allows us to clearly analyze the effect of variations in these properties on the geo-effectiveness of CME-CME interactions.

\subsubsection{Density} 
Figure \ref{fig:Dst_min} illustrates the minimum Dst values for low-density and high-density cases across three different longitudes: -10\textdegree, 0\textdegree, and +10\textdegree. The percentage differences (nearest integer of $100\cdot\left[\frac{HD-LD}{LD}\right]$) between paired cases highlight some significant trends. At 0\textdegree\ (panel a2) and +10\textdegree\ (panel a3), high-density cases consistently show lower minimum Dst values compared to low-density cases, with percentage differences ranging from 42\% to 70\% at 0\textdegree\ and 18\% to 129\% at +10\textdegree. This indicates that higher initial densities tend to exacerbate the severity of geomagnetic storms at these longitudes. This trend seems intuitive since higher initial density in CME2 leads to a stronger trailing shock and higher in-situ speed and density, which in turn would lead to a lower Dst index. However, at -10\textdegree\ (panel a1), the trend is reversed; high-density cases exhibit higher minimum Dst values compared to low-density cases, with percentage differences ranging from -8\% to -23\%. Although there is a significant increase in in-situ speed at -10\textdegree, the in-situ density does not show a major increment, possibly because the trailing shock has to cover a much larger radial distance in an over-expanded region of CME1. Additionally, when this second shock arrived, the B$_{\rm z}$  profile was not in southward direction as it was along the 0\textdegree\ and +10\textdegree\ longitudes. Due to this combined in-situ configuration, the effective Dst minimum became higher.

\subsubsection{Speed} 
At -10\textdegree\ (panel b1), the differences between low-speed and high-speed cases are minimal, with percentage changes ranging from -11\% to 2\%. This suggests that initial speed has a relatively minor impact on the severity of geomagnetic storms along this over-expanded CME1 region. In contrast, at 0\textdegree\ (panel b2), high-speed cases consistently exhibit lower minimum Dst values compared to low-speed cases, with percentage differences ranging from 7\% to 15\% and an average change of 11\%. 
This potentially indicates that higher initial speeds of CME2 results in a more geoeffective configuration at this longitude.
At +10\textdegree\ (panel b3), the trend is less consistent (in 6 out of 8 cases), with percentage differences varying from -2\% to 18\% and an average change of 5\%. Despite this variability, high-speed cases generally lead to lower minimum Dst values (in 18 out of 24 cases), suggesting a trend towards more severe storms with increased speeds. These findings demonstrate that while the influence of initial speed is significant at 0\textdegree\ and somewhat at +10\textdegree, it is relatively minor at -10\textdegree.

\subsubsection{Tilt} 
Like initial density and speed, the effect of CME2 tilt on minimum Dst values also shows distinct patterns across the three longitudes. With an average percentage difference of 3\%, ranging from -2\% to 6\%, the Dst minimum was lower for 6 out of 8 tilt cases along -10\textdegree\ longitude (panel c1). At 0\textdegree\ (panel c2), the tilt's effect is more varied, with percentage differences ranging from -5\% to 28\% and an average change of 7\%, showing lower Dst minimum values for tilted cases in 5 out of 8 instances. The most pronounced effect is observed at +10\textdegree\ (panel c3), where the differences are consistent and substantial, ranging from 32\% to 128\% with an average decrease of 81\%. This significant impact at +10\textdegree\ highlights how tilt in CME2, which alters the magnetic field orientation, can greatly intensify the geomagnetic storm along one direction. Ideally, the Dst minimum should have consistently increased in the opposite direction (+10\textdegree), but the non-uniform deformation of CME1 due to inhomogeneous ambient solar wind causes the Dst values to deviate from this expected trend.

\subsubsection{Magnetic Flux} 
The increase in the initial magnetic flux of a CME results in an increase in its magnetic field strength. In an ideal isolated flux rope scenario, this enhancement would typically lead to a lower Dst minimum upon the CME's interaction with Earth's magnetosphere. However, for CME-CME interactions in the presence of a realistic non-uniform ambient solar wind, the outcome can differ significantly. This is evident in the Figure \ref{fig:Dst_min}. At -10\textdegree\ (panel d1), the differences between low and high magnetic flux cases are relatively minor, with percentage changes ranging from 2\% to 12\% and an average difference of 5\%. This consistent trend of lower Dst minimum for higher initial magnetic flux suggests that magnetic flux does impact the severity of geomagnetic storms, albeit slightly at this location. At 0\textdegree\ (panel d2), the impact of magnetic flux is more pronounced, with percentage differences ranging from -2\% to 36\% and an average change of 13\%. The trend is consistent, with one exception that has the lowest percentage change. Combining the results from -10\textdegree\ and 0\textdegree, the trend is clear: higher magnetic flux leads to lower Dst minimum values in 15 out of 16 cases. However, at +10\textdegree\ (panel d3), the trend is reversed in 7 out of 8 cases; high magnetic flux cases exhibit higher minimum Dst values compared to low magnetic flux cases, with percentage differences ranging from -46\% to 2\% and an average increase of -16\%. It is important to emphasize that this trend reversal occurs along the longitude where the interaction has been strongest. This potentially indicates that higher magnetic flux in CME2 may lead to greater magnetic flux dissipation in strong CME-CME interaction regions, reducing the geo-effectiveness of CME-CME interactions at this longitude.

\subsection{Cumulative Dst} \label{sec: Dst_cum}

\begin{figure*}
       \centering
       \includegraphics[width = \columnwidth]{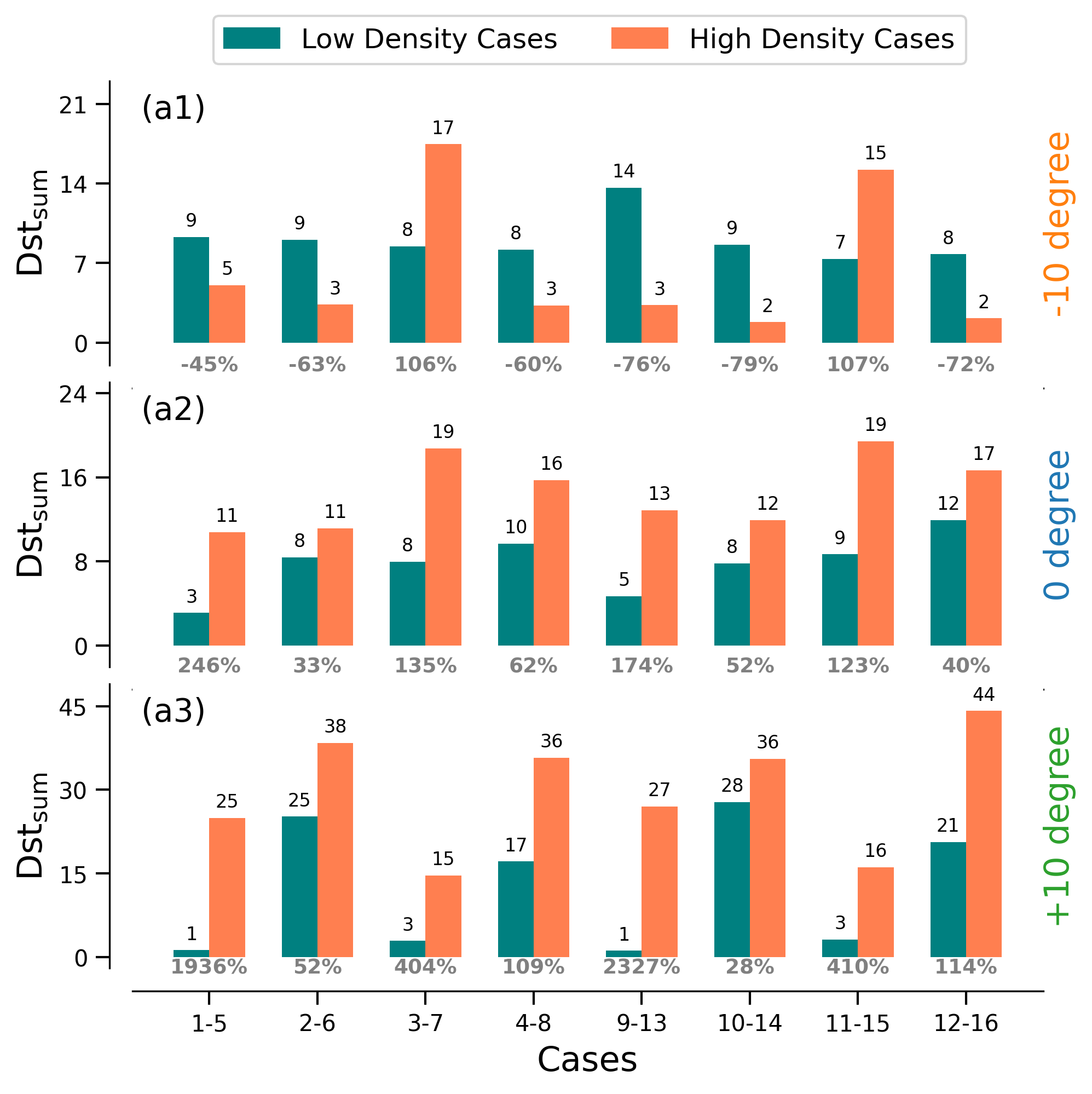}
       \includegraphics[width = \columnwidth]{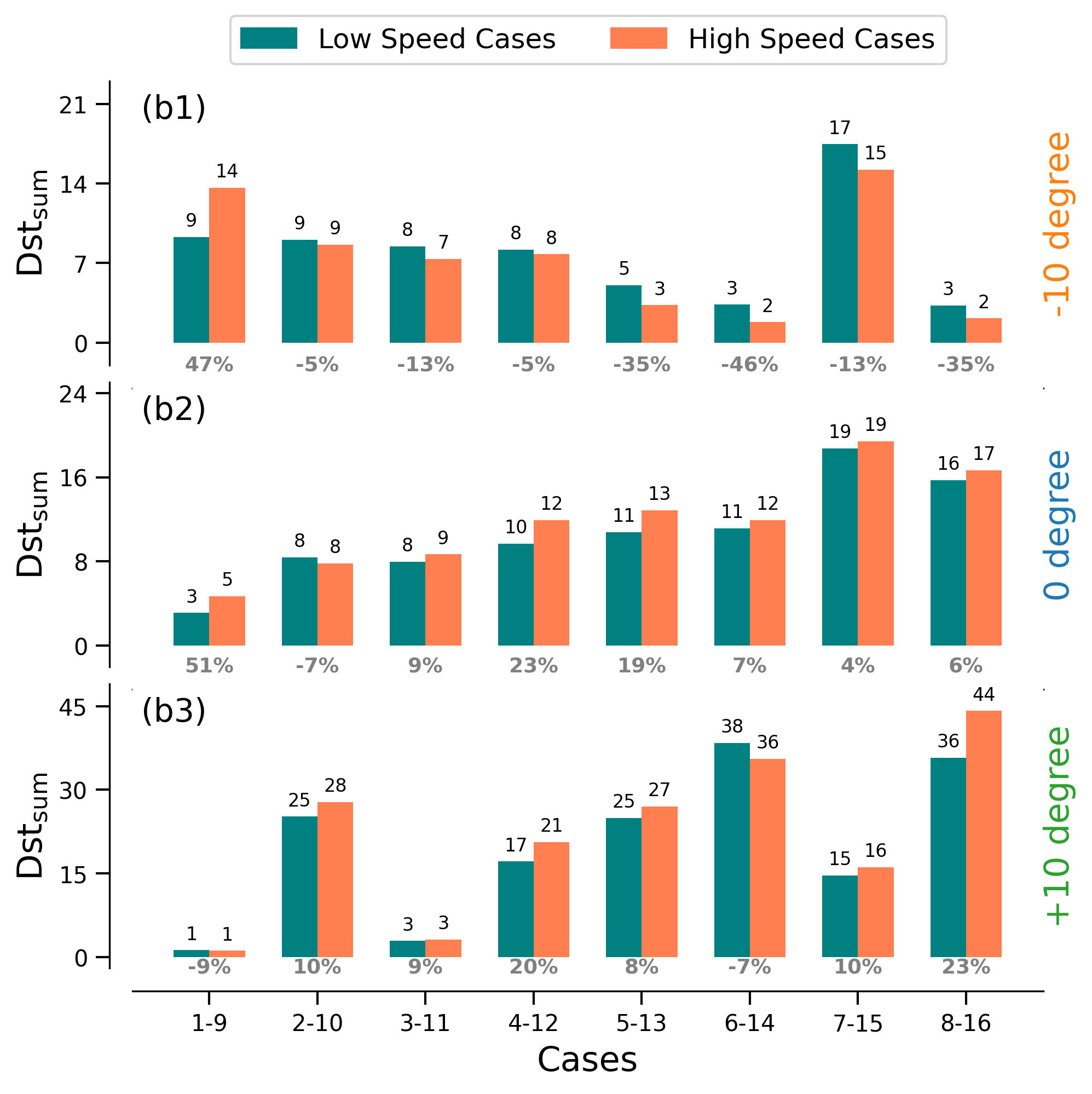}
       \includegraphics[width = \columnwidth]{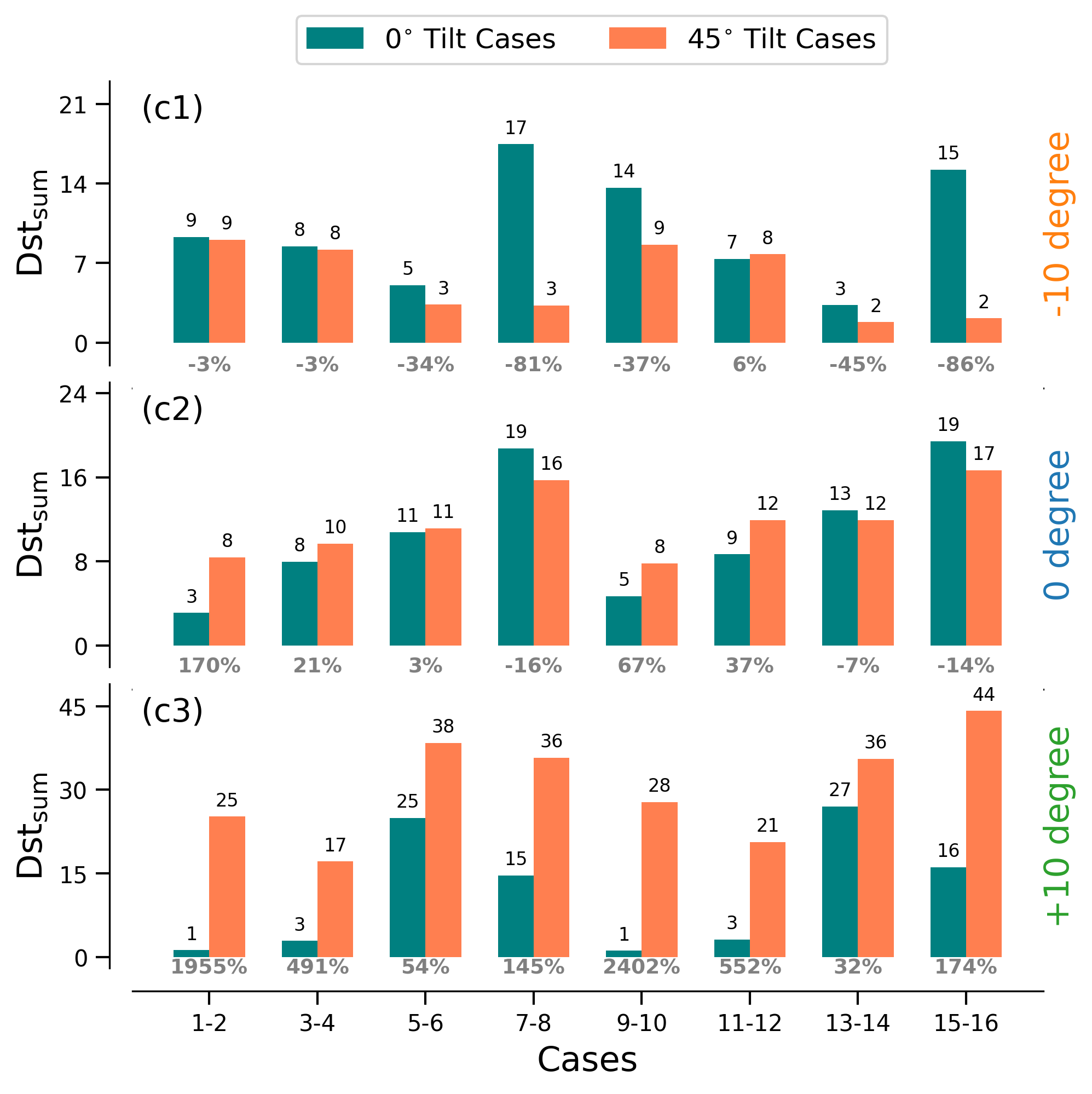}
       \includegraphics[width = \columnwidth]{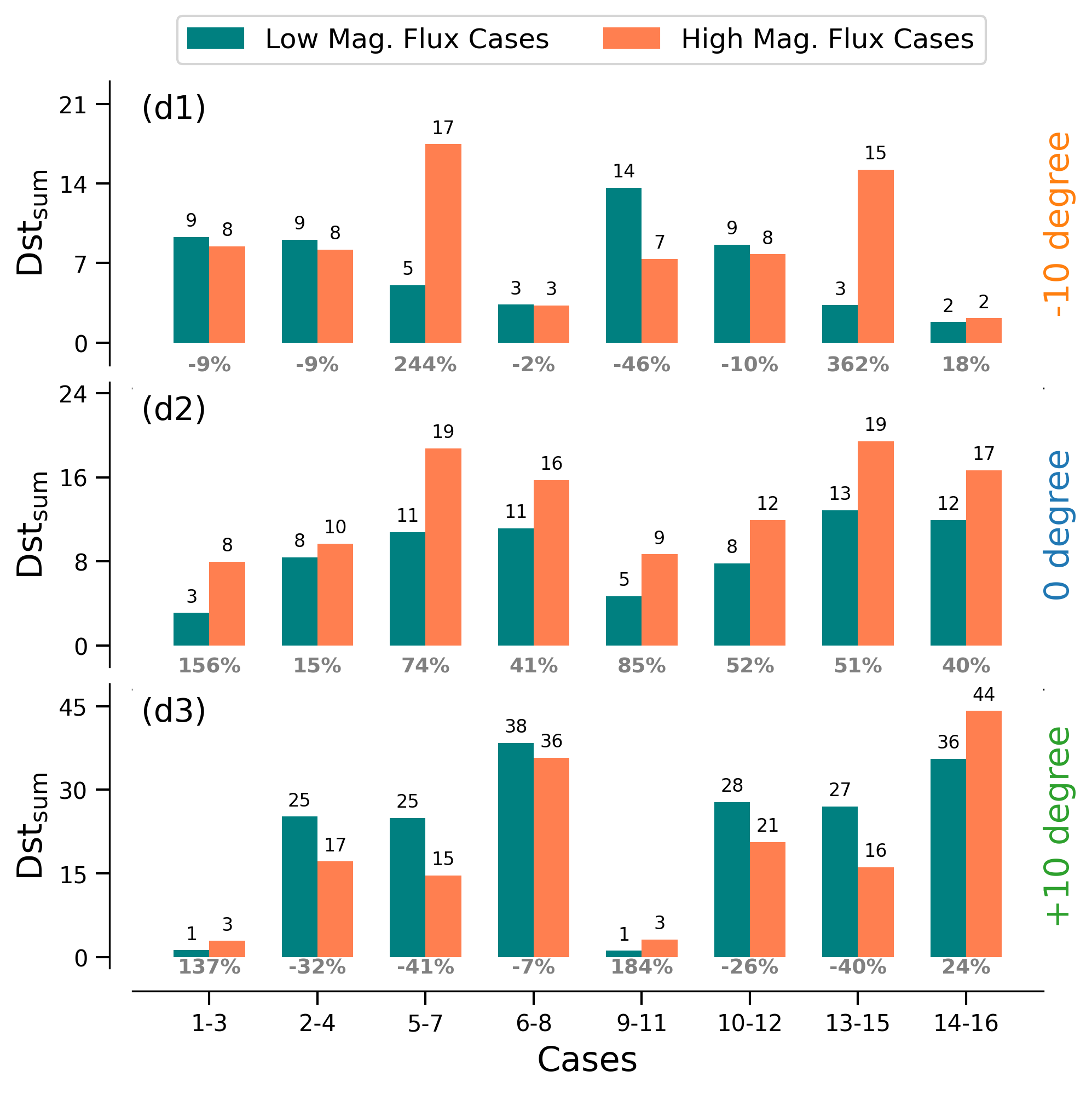}
       \caption{The histogram plots of the cumulative Dst values corresponding to the change in initial density (a1-a3), speed (b1-b3), tilt (c1-c3) and magnetic flux (d1-d3), at three longitudes at 1AU.}
       \label{fig:Dst_sum}
\end{figure*}

We use the cumulative Dst value, representing the summation of the Dst index until it returns to pre-storm levels, to assess the overall impact of the simulated geomagnetic storm event. 
By considering the cumulative impact and not just the minimum value of Dst we can compare the overall intensity of events with different duration \citep{lotz2015}.
Combined with the Dst minimum, this metric encapsulates both the intensity of the storm and its recovery characteristics, offering a more complete picture of the storm's effect on the Earth's magnetosphere. Similar to the previous subsection, we have used a pair-based analysis to identify trends in the effects of the initial properties of CME2.

\subsubsection{Density} 
The histograms in Figure \ref{fig:Dst_sum} illustrate the significant impact of initial density on the cumulative Dst values. Panel a1 shows the cumulative Dst index values for -10\textdegree\ longitude. The differences between low-density and high-density cases are substantial, with percentage changes ranging from -79\% to 107\% and an average change of -29\%. This trend of higher initial densities resulting in lower cumulative Dst values is observed in 6 out of 8 cases, indicating a general reduction in geomagnetic impact at this longitude, consistent with the trend observed in the minimum Dst. At 0\textdegree\ (panel a2), high-density cases exhibit consistently higher cumulative Dst values compared to low-density cases, with percentage differences ranging from 33\% to 246\% and an average increase of 109\%. At +10\textdegree\ (panel a3), the differences are even more pronounced, with percentage changes ranging from 28\% to 2327\% and an average increase of 731\%. This increasing trend is observed in all 8 cases at 0\textdegree\ and +10\textdegree, underscoring the substantial enhancement in storm duration due to higher initial densities of CME2. These findings demonstrate that the initial density of CME2 plays a crucial role in determining the cumulative geomagnetic impact, with the most significant effects observed at 0\textdegree\ and +10\textdegree, and a complex behavior at -10\textdegree.

\subsubsection{Speed} 
At -10\textdegree\ (panel b1), the cumulative Dst values show a wide range of percentage changes, from -46\% to 47\%, with an average change of -16\%. This variability highlights that higher initial speeds tend to reduce the cumulative geomagnetic impact at this location, though the trend is opposite to what was observed in the minimum Dst, and the effect is not uniform. In contrast, at 0\textdegree\ (panel b2), the cumulative Dst values for high-speed cases exceed those of low-speed cases, with percentage changes spanning from -7\% to 51\% and an average increase of 16\%. Here, 7 out of 8 cases follow this pattern, consistent with the trend observed in the minimum Dst. At +10\textdegree\ (panel b3), the differences are more subtle, with percentage changes ranging from -9\% to 23\% and an average change of 9\%. This pattern appears in 6 out of 8 cases, suggesting a mild tendency for higher speeds to amplify the cumulative Dst values. These observations indicate that the impact of CME2's initial speed on storm duration and intensity varies with longitude, showing enhancement at 0\textdegree\ and +10\textdegree, but reduction at -10\textdegree.

\subsubsection{Tilt} 
The impact of tilt between CMEs shows a relatively more consistent trend. For -10\textdegree, the cumulative Dst values decrease, show mixed behavior at 0\textdegree, and increase at +10\textdegree. At -10\textdegree\ (panel c1), the high tilt cases have percentage changes ranging from -86\% to 6\%. This trend is observed in 7 out of 8 cases, with an average change of -36\%, opposed to the general observation in minimum Dst. Moving to 0\textdegree\ (panel c2), the effect of tilt becomes inconsistent. Here, percentage changes vary from -16\% to 170\%, with an average change of 34\%. The trend is followed in 5 out of 8 cases, indicating a mixed influence of tilt on cumulative geomagnetic impact, and is not as pronounced as at -10\textdegree. In contrast, at +10\textdegree\ (panel c3), the high tilt cases dramatically increase the cumulative Dst values, with percentage changes ranging from 32\% to 2402\% and an average increase of 697\%. This overwhelming trend is observed in all 8 cases, highlighting a substantial enhancement in geomagnetic storm duration and intensity due to tilt at this longitude. This demonstrates that CME2’s tilt has a variable influence on the cumulative geomagnetic impact, with a significant reduction at -10\textdegree, moderate variability at 0\textdegree, and immense enhancement at +10\textdegree.

\subsubsection{Magnetic Flux} 
Among the three longitudes, only the cumulative Dst values at 0\textdegree\ show a consistent trend, while the $\pm$10\textdegree\ longitudes exhibit mixed behavior with increasing magnetic flux of CME2. This inconsistency contrasts with the statistical trends observed in the minimum Dst. At -10\textdegree\ (panel d1), percentage changes range from -46\% to 362\%. Decreases are observed in 5 out of 8 cases, but the average change is +56\%, reflecting minor decrements and significant increments. Similarly, at +10\textdegree\ (panel d3), the results are highly variable, with percentage changes ranging from -41\% to 184\% and an average increase of 22\%. Decreases occur in 5 cases, while increases are seen in 3 out of 8 cases. In contrast, at 0\textdegree\ (panel d2), the impact of higher magnetic flux is consistent and more pronounced, with percentage changes spanning from 15\% to 156\% and an average increase of 67\%. This trend is observed in all 8 cases, indicating a strong correlation between increased magnetic flux and enhanced geomagnetic storm duration and intensity.

\section{Summary} \label{sec:summary}

In this ensemble study, we utilized the SWASTi framework to conduct MHD simulations of 16 different scenarios of CME-CME interaction. The simulations employed a flux rope CME, propagating within data-driven realistic solar wind conditions. The chosen solar wind conditions correspond to Carrington rotation period 2270, with the projected trajectory of the CMEs passing through a solar wind stream interaction region (SIR). This unique setup implies that all CME-CME interactions in this study are significantly influenced by the in-path SIR, effectively making it a CME-CME-SIR interaction scenario. Below are brief discussions and conclusive remarks on the topics covered in this work:

\textit{\textbf{Role of Solar Wind}}: Figure \ref{fig:2D_example} clearly illustrates this situation and its implications. In all ensemble cases, the upper flank of CME1 over-expanded compared to the lower flank. This differential expansion occurs because the upper flank, situated along the fast stream, experiences greater pressure gradients, leading to more expansion and faster movement than the bottom flank. The inhomogeneity in the ambient solar wind results in an asymmetrical radial width of CME1. This variability makes the CME-CME collision non-uniform across the interaction surface.

\textit{\textbf{Shock Evolution}}: We thoroughly investigated the dependency of shock evolutionary stages proposed by \cite{nolugaz_2005_numerical} on the initial properties of CME2. Our study revealed that the evolution of these stages varied across different longitudes, primarily due to the non-uniform radial extension of CME1. This finding highlights that the shock-based classification of CME-CME interaction stages is not universally applicable to the entire structure but rather a localized phenomenon.

\textit{\textbf{Impact on CME1}}: By comparing ensemble cases with the single CME1 simulation, we found some peculiar trends in the kinematic, magnetic, and structural properties: 

\begin{enumerate}[label={\tiny \textbullet}, leftmargin=*, itemindent=+0.75em]

    \item Kinematic: Due to the collision, CME1 gained 9-36\% in radial momentum and 15-65\% in kinetic energy by the time it reached 1 AU. Their temporal evolution showed similar patterns, with an initial rising phase lasting about 10 hours, followed by a diminishing phase. The amount of gain and the duration of the rising phase, were most influenced by CME2's initial density.
    
    \item Magnetic: CME1's magnetic properties showed less significant changes compared to its kinematic properties and followed a different evolution pattern. It increased by 1-3\% only in high-density cases. We observed an initial increase in magnetic energy up to 20-25 hours, followed by a decrease due to magnetic field dissipation. This behavior is consistent with \cite{koehn_2022_successive}, given that the CMEs had the same chirality.

    \item Structural: The longitude (+10\textdegree), where CME1 interacted with both the SIR from the front and CME2 from behind, showed the highest compression -- more than twice at longitude (-10\textdegree) without SIR interaction. This indicates that the SIR significantly enhances the leading CME's radial compression. Overall, CME2's initial density was observed to be a key factor in determining the interaction's impact on CME1.
    
\end{enumerate}

\textit{\textbf{Mixing}}: We also analyzed the mixing of CMEs during their interaction in the heliosphere and found that the amount of mixing varies significantly depending on the initial conditions of the trailing CME. Enhanced mixing was observed along the bottom flank, where the interaction strength is higher due to the presence of the SIR ahead of the leading CME, again highlighting the impact of inhomogeneity in the ambient solar wind.

\textit{\textbf{Reverse Shock}}: Another noteworthy phenomenon observed in the CME-CME interaction was the formation of reverse shocks in cases of strong interaction. Similar observations were reported by \cite{nolugaz_2005_numerical} in their MHD simulation of two identical CMEs interacting in a simple axis-symmetric solar wind setup. Additionally, \cite{trotta_2024_observation} recently observed such a reverse shock using Solar Orbiter data. We found that these shocks can originate from multiple locations due to the non-uniform interaction between CME1 and CME2 along different longitudes. As these reverse shocks propagate inside CME2, they create a complex pattern of alternating compressed and rarefied regions, causing ripples or fluctuations in the in-situ data (see Figure \ref{fig:shock_duration}). These ripples influence the geo-effectiveness of the CME-CME structure, notably extending the overall recovery phase and delaying the return to pre-storm Dst levels.

\textit{\textbf{Geo-effectiveness}}: We conducted a statistical study of the minimum and cumulative Dst index values for the ensemble cases to identify trends related to CME2's initial properties. Along the strong interaction region (+10\textdegree), the minimum Dst value decreased with increased initial density, tilt, and speed of CME2, with average changes of 66\%, 81\%, and 6\%, respectively, indicating higher geo-effectiveness. Conversely, the minimum Dst value increased by an average of 19\% with higher initial magnetic energy, suggesting greater magnetic dissipation and lower geo-effectiveness. Trends were less consistent at 0\textdegree\ and -10\textdegree. Overall, increase in CME2's any initial property mostly leads to stronger (72\% of cases) and prolonged (63\% of cases) storm.

It is important to emphasize that this ensemble study was conducted using an ideal MHD setup. Consequently, a quantitative study of magnetic dissipation in CME-CME interactions lies beyond the scope of this work. Additionally, the ambient solar wind conditions were consistent across all cases, introducing a bias towards specific ambient conditions. Therefore, the conclusions drawn are particularly relevant to the scenario of CME-CME-SIR interaction. We have also not considered the role of initial flux rope orientation (e.g., chirality and polarity) in this study, which can have considerable impact on geo-effectiveness \citep{koehn_2022_successive}. Moreover, the CME insertion method used here is not fully conventional, and a more organic leg-cutting approach is needed to improve simulation fidelity. In our future work, we aim to include non-ideal MHD effects to explore in greater depth, especially to investigate the formation of reverse shocks and their dependency on CME properties. This approach will provide a more comprehensive understanding of the underlying physical processes that can enhance the geo-effectiveness of CME-CME interaction events.

\begin{acknowledgments}
We thank the anonymous referee for valuable suggestions, which enhanced the content of this manuscript.
PM gratefully acknowledges the support provided by the Prime Minister's Research Fellowship. Much of the work in this paper was conducted while PM was hosted by SL at SANSA in Hermanus (South Africa), funded by the SCOSTEP Visiting Scholar (SVS) programme. We are grateful for funding from the SVS programme and hosting by SANSA. BV and DC acknowledge the support from the ISRO RESPOND grant number: ISRO/RES/2/436/21-22. Furthermore, the work of DC is supported by the Department of Space, Government of India.
\par
The used GONG synoptic magnetograms maps can be freely obtained from \url{https://gong.nso.edu/data/magmap/crmap.html}. The OMNI data are taken from the Goddard Space Flight Center, accessible at \url{https://spdf.gsfc.nasa.gov/pub/data/omni/}. The PLUTO code used for MHD simulation can be downloaded free of charge from \url{http://plutocode.ph.unito.it/}.
\end{acknowledgments}

\newpage

\appendix

\section{Empirical Relation of Dst} \label{sec: Dst_relation}

\cite{obrien_2000_forecasting} proposed the following empirical relations to compute the Dst index based on the plasma properties at Sun-Earth L1 point.

    \begin{equation}\label{eq:Dst1}
       \frac{dDst^*}{dt} = Q(t) - \frac{Dst^*}{\tau}  ,
    \end{equation}
    \begin{equation}\label{eq:Dst2}
        Q = -4.4(VB_{S} - 0.5)  ,
    \end{equation}
    \begin{equation}\label{eq:Dst3}
        \tau = 2.4 \exp\left(\frac{9.74}{4.69 + VB_{S}}\right)  ,
    \end{equation}
    \begin{equation}\label{eq:Dst4}
        VB_{S} = 
                \left\{
                \begin{array}{ll}
                |VB_{z}| & \text{if } B_z < 0  , \\
                0 & \text{if } B_z \geq 0  ,
                \end{array}
                \right.
    \end{equation}
    \begin{equation}\label{eq:Dst5}
        Dst = Dst^* + b\sqrt{P_{dyn}} - c  .
    \end{equation}

Here, Dst$^*$ represents the pressure corrected Dst index, accounting for magnetopause current contamination, with constants $b = 7.26$ and $c = 11$. The parameter $Q$ denotes the rate of energy injection into the ring current, while $\tau$ represents the decay time of the ring current, influenced by particle loss to the atmosphere. The variables $V$ and $P_{dyn}$ are the plasma speed and dynamic pressure, respectively, and $B_z$ is the Z-component of the magnetic field in geocentric solar magnetospheric (GSM) coordinates. Dst$^*$ is used to perform time integration, and the model output Dst is calculated from Equation \ref{eq:Dst5}. 
Under usual circumstances the initial value of the estimated Dst is set to the last measured value before the prediction is made \citep{obrien_2000_forecasting}. 
However, since we are dealing with simulated CMEs we can't have observed initial values for the Dst. 
Therefore we set the ``initial level'' of Dst at 0 nT, as Dst was designed for the quiet time reference level to be zero \citep{Sugiura}.

Although the efficiency of the above empirical Dst relation has been extensively demonstrated by \cite{obrien_2000_forecasting}, we sought to verify its performance for significant geoeffective events in recent years. To test this relation, we compared the Dst values derived from the model with OMNI 1-hour data. Specifically, we examined CME events occurring during Carrington rotations (CR) 2165, 2194, and 2270, which included five interacting CMEs, two interacting CMEs, and one single CME, respectively. For these comparisons, the initial Dst values required for the model were set to match the observed Dst values at the initial time for CR.

\begin{figure}
           \centering
           \includegraphics[width = \columnwidth]{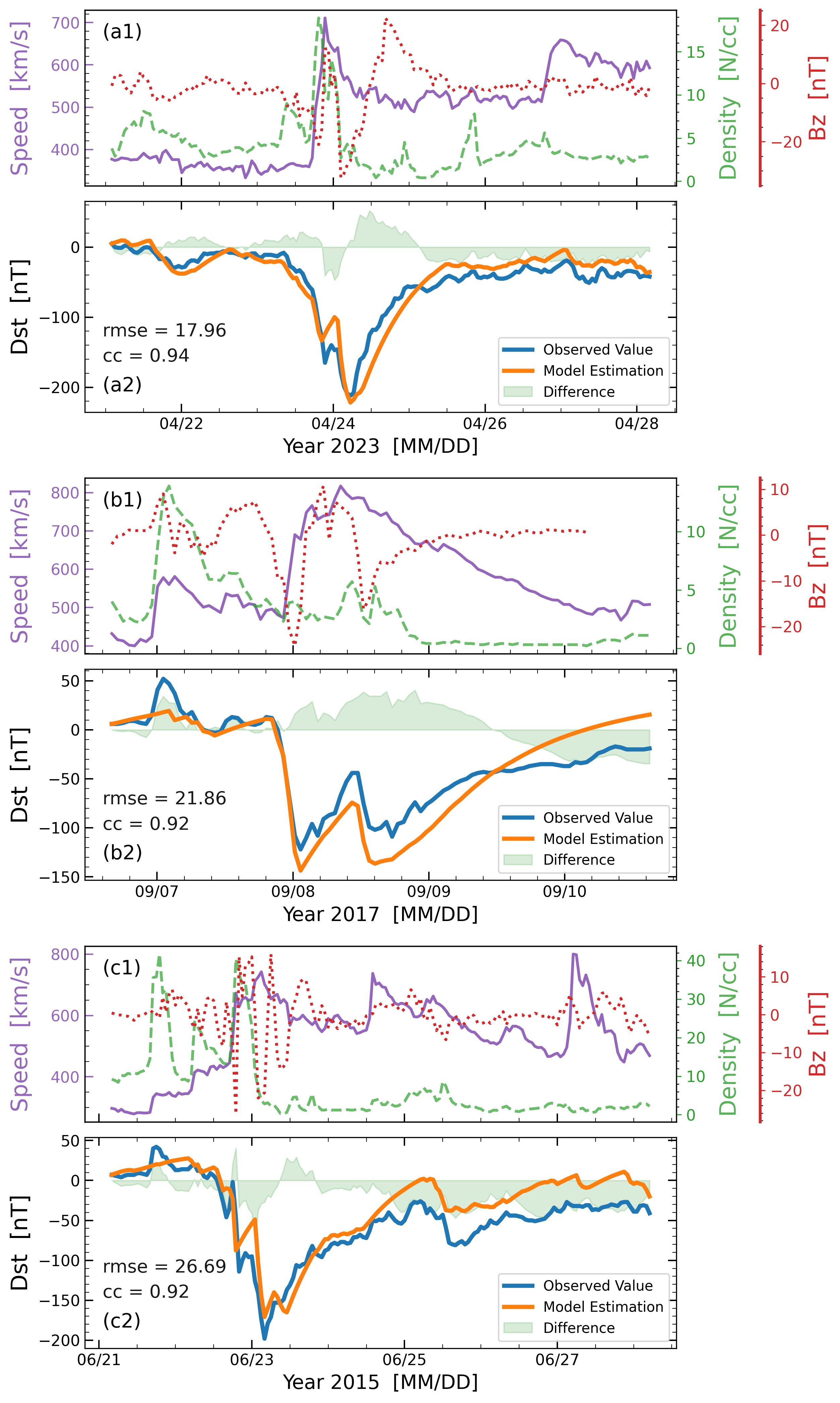}
           \caption{Comparison of observed Dst and the estimated values as described in equations (\ref{eq:Dst1})-(\ref{eq:Dst5}). The a1, b1 \& c1 panels show the observed in-situ solar wind speed, density, and B$_{\rm z}$ values at Sun-Earth L1 point, for CR2270, CR2194 and CR2165, respectively. The a2, b2 \& c2 panels show the observed versus estimated Dst values for these three CRs, along with their difference.}
           \label{fig:Dst_test}
    \end{figure}

Figure \ref{fig:Dst_test} presents the comparison between the modeled and observed Dst values. Subplots (a1) and (a2) depict the observed in-situ plasma properties at L1 (speed, density, and IMF B$_{\rm z}$ component). The estimated Dst, based on these values, is broadly accurate and captures most of the storm's features, such as the momentary rise in Dst due to a short-period northward IMF on June 1, 2023 (see Figure \ref{fig:Dst_test} a3). The onset of the initial phase of Dst and the duration of the main phase in the model output also match nicely for all three CRs. However, some finer structures are missed, such as the sudden commencement on August 18, 2017 (panel b), and the model shows some discrepancies in the magnitude of the storm's recovery phase.

The statistical results are promising, with Pearson correlation coefficient (cc) values ranging from 0.92 to 0.94 and root mean square error (rmse) values between 18 and 27 nT, corresponding to an error margin of roughly 9 to 17\%. This strong agreement indicates that the empirical relation is reliable for comparative studies between events, which is the primary application in this work for comparing the Dst values across different ensemble cases.

\section{Interaction Scenario}
\begin{figure*}
       \centering
       \includegraphics[width = 0.9\textwidth]{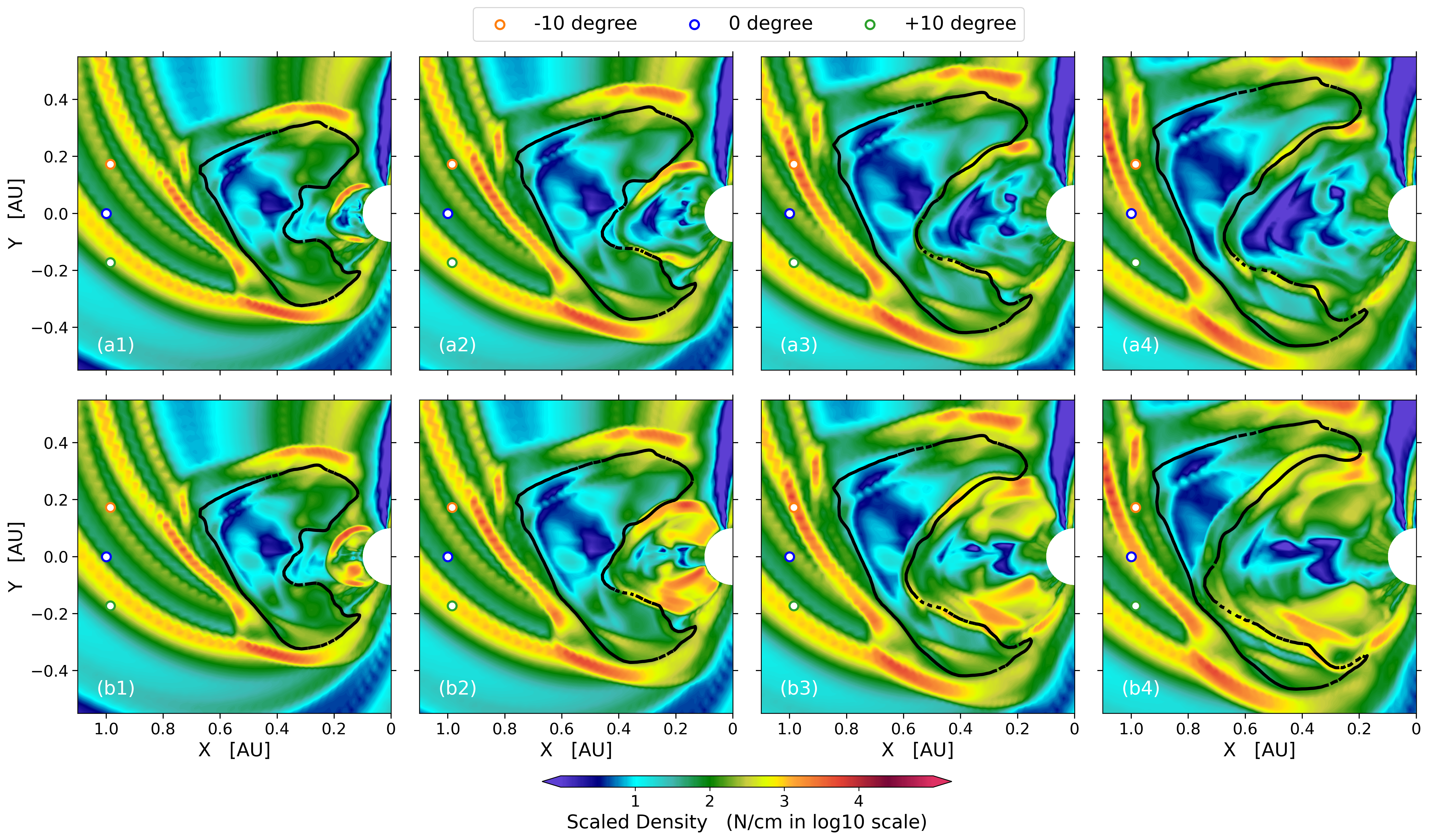}
       \caption{The panels showcases the scaled density (in log10 scale, N/cm³) for the considered CME-CME-SIR interaction scenario in the inner-heliosphere. Subplots (a1)-(a4) represent the temporal evolution of scale density for case LSLDLF0, while subplots (b1)-(b4) depict the scenario for case LSHDLF0.}
       \label{fig:rho_shock}
    \end{figure*}

The nature of CME-CME interaction scenario undertaken in this study is demonstrated in Figure \ref{fig:rho_shock}. It showcases a unique interaction scenario complicated by the presence of SIR ahead of the leading CME. In the subplots, the trailing CME is catching up to the leading CME, which has already encountered the SIR. This configuration creates a dynamic environment where multiple structures — two CMEs and the SIR — interact, leading to a complex scenario. High-density regions at the boundaries of both CMEs and within the SIR are clearly visible. This particular scenario underscores the complexity of interacting solar wind structures, with the scaled density plots offering a detailed view of the varying intensities within the system. The corresponding temperature plot has been shown in Figure \ref{fig:2D_shock}.

\section{In-situ profile at 1AU}

\begin{figure*}[b]
       \centering
       \includegraphics[width = 0.9\textwidth]{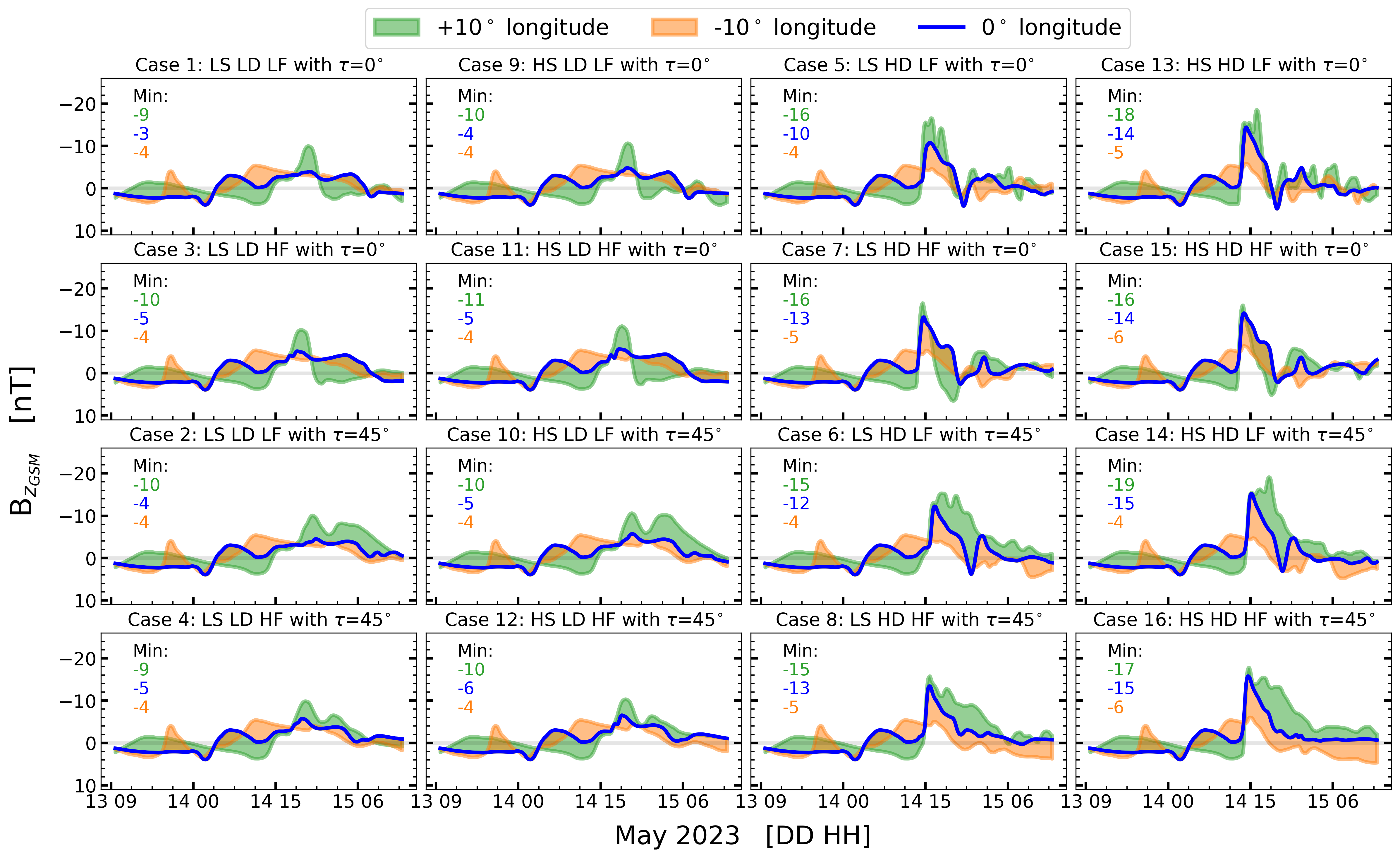}
       \caption{These plots illustrate the southward magnetic field (B$_{\rm z}$) at the location of three virtual spacecrafts used in this study, showcased at -10 (orange), 0 (blue), and +10 (green) degree longitudinal positions. The entire set of 16 cases within the CME-CME interaction ensemble is represented here.}
       \label{fig:Bz_all}
    \end{figure*}

In this section, we present the in-situ profiles observed by three virtual spacecraft at 1 AU along 0\textdegree\ and $\pm$10\textdegree\ longitudes for all the 16 ensemble cases. Figure \ref{fig:Bz_all} shows the density and southward component of the magnetic field (B$_{\rm z}$) profiles, for the duration from 13th May 9 UT to 15th May 14 UT, 2023, to highlight the most relevant structures. To keep the relevant peaks in the upper half of the subplots, we have inverted the B$_{\rm z}$ profile, with negative values shown upward and positive values downward. The key aspects of the features in these in-situ profiles have been discussed in Section \ref{sec: dst_variation}.

\bibliography{CME-CME}{}
\bibliographystyle{aasjournal}

\end{document}